\documentclass[article,twocolumn,prl,superscriptaddress,tightenlines,nofootinbib]{revtex4}

\usepackage{amssymb}
\usepackage{enumerate}
\usepackage{amssymb}
\usepackage{amsmath}
\usepackage{tikz}
\usepackage[compat=1.1.0]{tikz-feynman}
\usepackage{feynmf}
\usepackage{makecell}
\usepackage{slashed}
\usepackage{natbib}
\usepackage{graphicx}
\usepackage[pdftex,bookmarks,linktocpage,pdfpagelabels,plainpages=false,hyperfigures,linkcolor=blue,citecolor=blue]{hyperref} 
\hypersetup{colorlinks=true}

\usepackage{mathrsfs,amssymb}  
\usepackage{cancel}
\usepackage[normalem]{ulem}

\newcommand\be{\begin{equation}}
\newcommand\ee{\end{equation}}
\newcommand{\comment}[1]{}
\newcommand\bea{\begin{eqnarray}}
\newcommand\eea{\end{eqnarray}}

\begin{document}

\begin{flushleft}
MI-TH-1942
\end{flushleft}
\bibliographystyle{apsrev4-1}

\title{New Directions for Axion Searches via Scattering at Reactor Neutrino Experiments}

\author{James B.~Dent} 
\affiliation{Department of Physics, Sam Houston State University, Huntsville, TX 77341, USA}

\author{Bhaskar Dutta}
\affiliation{Mitchell Institute for Fundamental Physics and Astronomy,
   Department of Physics and Astronomy, Texas A\&M University, College Station, TX 77845, USA}

\author{Doojin Kim}
\affiliation{Mitchell Institute for Fundamental Physics and Astronomy,
   Department of Physics and Astronomy, Texas A\&M University, College Station, TX 77845, USA}

\author{Shu Liao}
\affiliation{Mitchell Institute for Fundamental Physics and Astronomy,
   Department of Physics and Astronomy, Texas A\&M University, College Station, TX 77845, USA}

\author{Rupak Mahapatra}
\affiliation{Mitchell Institute for Fundamental Physics and Astronomy,
   Department of Physics and Astronomy, Texas A\&M University, College Station, TX 77845, USA}
   
\author{Kuver Sinha}
\affiliation{Department of Physics and Astronomy, University of Oklahoma, Norman, OK 73019, USA}

\author{Adrian Thompson}
\affiliation{Mitchell Institute for Fundamental Physics and Astronomy,
   Department of Physics and Astronomy, Texas A\&M University, College Station, TX 77845, USA}

\begin{abstract}

Searches for pseudoscalar axion-like-particles (ALPs) typically rely on their decay in beam dumps or their conversion into photons in haloscopes and helioscopes. We point out a new experimental direction for ALP probes via their production by the intense gamma ray flux available from megawatt-scale nuclear reactors at neutrino experiments through Primakoff-like or Compton-like channels. Low-threshold detectors in close proximity to the core will have visibility to ALP decays and inverse Primakoff and Compton scattering, providing sensitivity to the ALP-photon and ALP-electron couplings. We find that the sensitivity to these couplings at the ongoing MINER  and  various other reactor based neutrino  experiments, e.g., CONNIE, CONUS, $\nu$-cleus etc. exceeds existing limits set by laboratory experiments and, for the ALP-electron coupling, we forecast the world's best laboratory-based constraints over a large portion of the sub-MeV ALP mass range.
\end{abstract}

\maketitle

\noindent {\bf{\emph{Introduction. - }}}Axions are a well-motivated and extensively explored extension of the Standard Model (SM) both for their ability to address the strong CP problem \cite{Peccei:1977hh,Wilczek:1977pj,Weinberg:1977ma}, and for serving as a dark-matter candidate (see, for example, the reviews \cite{Duffy:2009ig,Marsh:2015xka,Battaglieri:2017aum}). Theoretical studies have branched away from solely investigating the original QCD axion, the pseudoscalar which can solve the strong CP problem, and have incorporated general axion-like particles (ALPs) into a range of models.

Axions and ALPs are undergoing a period of intense experimental scrutiny from a wide array of approaches that exploit an axion-photon coupling in helioscopes such as CAST \cite{Zioutas:1998cc,Anastassopoulos:2017ftl} and IAXO \cite{Irastorza:2013dav}, haloscopes including Abracadabra \cite{Kahn:2016aff,Salemi:2019xgl}, ADMX \cite{Asztalos:2001tf,Du:2018uak}, CASPEr \cite{JacksonKimball:2017elr}, HAYSTAC \cite{Brubaker:2016ktl,Droster:2019fur}, light-shining-through-walls experiments including ALPSII \cite{Spector:2019ooq}, and additional experiments that exploit the possible axion-photon coupling through interferometry \cite{Melissinos:2008vn,DeRocco:2018jwe} such as ADBC \cite{Liu:2018icu} and DANCE \cite{Obata:2018vvr}. 
Additionally there are a variety of current and proposed beam dump and fixed target experiments that can search for $a\rightarrow\gamma\gamma$ decays or axion bremsstrahlung from electrons including FASER \cite{Feng:2018noy}, LDMX \cite{Berlin:2018bsc,Akesson:2018vlm}, NA62 \cite{Volpe:2019nzt}, SeaQuest \cite{Berlin:2018pwi}, and SHiP \cite{Alekhin:2015byh}. For a recent review of the current status and future prospects of axion searches at collider see, for example, Ref.~\cite{Bauer:2018uxu}. 
Neutrino experiments such as NOMAD~\cite{Astier:2000gx} have been used as ALP searches, and there are proposals such as PASSAT~\cite{Bonivento:2019sri} which are hybrids of the beam dump and helioscope approaches. 
Dark-matter direct detection experiments including XMASS~\cite{Abe:2012ut}, EDELWEISS-III~\cite{Armengaud:2018cuy}, LUX, \cite{Akerib:2017uem}, PandaX-II~\cite{Fu:2017lfc}, Xenon1T~\cite{Aprile:2019xxb}, and SuperCDMS~\cite{Aralis:2019nfa}, which have excellent electron recoil measurement capabilities, have also been used to search for ALP-electron scattering in addition to proposals for constraining this coupling through the use of geoscopes \cite{Davoudiasl:2009fe}. Direct detection experiments such as DAMA~\cite{Bernabei:2001ny}, EDELWEISS-II \cite{Armengaud:2013rta}, and XMASS \cite{Oka:2017rnn} also have demonstrated sensitivity to axion-photon couplings. Solar axions produced through nuclear transitions can also be searched for through resonant absorption by laboratory nuclei, which provides a bound on axion-nuclon couplings \cite{Moriyama:1995bz,Krcmar:1998xn,Krcmar:2001si,Derbin:2009jw,Gavrilyuk:2018jdi,Creswick:2018stb} (see also the brief discussions in the reviews \cite{Irastorza:2013dav,Sikivie:2020zpn}).

An ALP field $a$ could couple to SM particles through a myriad of operators, but the focus of this work is those of dimension-five, coupling $a$ to the electromagnetic current and its dual $g_{a\gamma\gamma}aF_{\mu\nu}\tilde{F}^{\mu\nu}$,
as well as the dimension-four operator  $g_{aee}a\bar{\psi}\gamma^5\psi$, coupling $a$ to electrons. 

In this paper, we focus on a new direction in ALP searches involving low-energy detectors at nuclear reactor facilities that will exploit both the copious photon production (and therefore, possible ALP production) and low-energy capabilities of the current detector technology. Specifically, we discuss the capabilities for probing ALPs at the upcoming search for coherent elastic neutrino-nucleus scattering (CE$\nu$NS) by the Mitchell Institute Neutrino Experiment at Reactor (MINER) Collaboration \cite{Agnolet:2016zir}, and at a few other reactor based CE$\nu$NS experiments, e.g., CONNIE~\cite{Aguilar-Arevalo:2016khx}, CONUS~\cite{Buck:2020opf}, and $\nu$-cleus~\cite{Strauss:2017cuu} experiments.
The MINER experiment consists of an array of low-threshold cryogenic germanium detectors sited a few meters from the core of the 1 MW nuclear reactor at the Nuclear Science Center 14 (NSC) at Texas A\&M University. 
The CONUS, CONNIE and $\nu$-cleus use GW reactors and Ge, Si-Skipper and CaWO$_4$(Al$_2$O$_3$) detector technologies, respectively. 
The nuclear reactor cores at these experiments will produce a copious amount of photons which can scatter off the material within the reactor tank to produce ALPs. 
On the detection side, the ALPs can directly scatter off detector nuclei and electrons, as well as decay in flight to photon or electron-positron pairs, providing a constraint on either the ALP-photon or ALP-electron coupling, respectively. 

In a previous reactor-based investigation by the TEXONO Collaboration~\cite{Chang:2006ug}, ALP production was modeled as arising from neutron capture or nuclear de-excitation with a branching ratio to ALPs (relative to photon production) that depends on the ALP mass, $m_a$, through an ALP-nucleon coupling, thus leading to weakening constraints as $m_a$ decreases. In the present work, however, we adopt a minimal approach where no ALP-nucleon coupling is assumed, and ALPs are produced via photon-induced scattering processes. These produce $m_a$-independent bounds for $m_a\lesssim 0.1$ MeV, allowing for broader coverage of the parameter space. Future work will consider inclusion of the nucleon coupling, which can improve sensitivity in some regions of parameter space.

We demonstrate that the current germanium configuration for MINER along with the ongoing CONUS, CONNIE and $\nu$-cleus experiments can 
become the most sensitive laboratory-based detectors to $g_{a\gamma\gamma}$ within an ALP mass range of $\sim(1 - 10^6)$ eV, and gain access to a wide swath of new parameter space in a similar mass range over several orders of magnitude in the coupling $g_{aee}$. These results speak to the tremendous opportunity for low-threshold detectors at nuclear reactor facilities and/or CE$\nu$NS experiments to search for ALPs.

In this paper we focus on laboratory searches which control both the production and detection sectors of the ALP processes. The ALP parameter space is also investigated by the astrophysical processes, however,  there exist several possible ways to circumvent astrophysical bounds that would exclude such a particle. These mechanisms have been discussed in the context of specific particle physics models, e.g., Refs.~\cite{Jaeckel:2006xm,Khoury:2003aq,Masso:2005ym,Masso:2006gc,Dupays:2006dp, Mohapatra:2006pv,Brax:2007ak}. 
These works investigate the environmental dependence of $m_a$ and $g_{a\gamma\gamma}$ which could allow the evasion of the bounds emerging from the null observation of ALPs at CAST ($m_a<0.01$~eV) and studies of the evolution of populations of RG and HB stars in globular clusters ($m_a<1$ keV).
For example, in Ref.~\cite{Mohapatra:2006pv}, a few additional scalars are introduced and the scalar dynamics are designed to invoke a phase transition below typical star temperatures. Consequently, the axion is produced along with a scalar whose mass may be above the stellar temperature ($\mathcal{O}(10)$ MeV). If this is the case, then axion production becomes exponentially suppressed such that astrophysical bounds would become irrelevant for any combination of $m_a$ and $g_{a\gamma\gamma}$ values projected to be constrained in this paper. 
In Refs.~\cite{Khoury:2003aq,Brax:2007ak} the axion is considered to be a chameleon-type field with its mass depending on the environmental matter density $\rho$. 
In this scenario, the axion effectively becomes much heavier inside stars so that the axion with $m_a\geq 10^{-1}$~eV (as measured in  the laboratory) does not suffer from the stellar constraints.  
Furthermore, Refs.~\cite{Masso:2005ym,Masso:2006gc,Dupays:2006dp} explored the possibility that the axion is a composite particle with a form factor leading to a suppression of the production in the stellar media which would evade the stellar constraints. 
In addition, they also considered models with a paraphoton where the ALPs are trapped in the stellar interior and they cannot freely escape, thus evading the stellar bounds.

\noindent {\bf \emph{ALP Production and Detection. - }}
We will focus on a generic model where the ALP can couple to either a photon or an electron as described by interaction terms in the Lagrangian of the form
\begin{equation}
\mathcal{L}_{\rm int} \supset -\frac{1}{4}g_{a\gamma\gamma}aF_{\mu\nu}\tilde{F}^{\mu\nu}-g_{aee}a\bar{\psi}_e\gamma_5\psi_e
\end{equation}
where $F_{\mu\nu}$ is the electromagnetic field strength tensor and its dual $\tilde{F}^{\mu\nu}=\epsilon^{\mu\nu\rho\sigma}F_{\rho\sigma}$.

\begin{figure*}[t]
    \centering
    \includegraphics[width=0.75\textwidth]{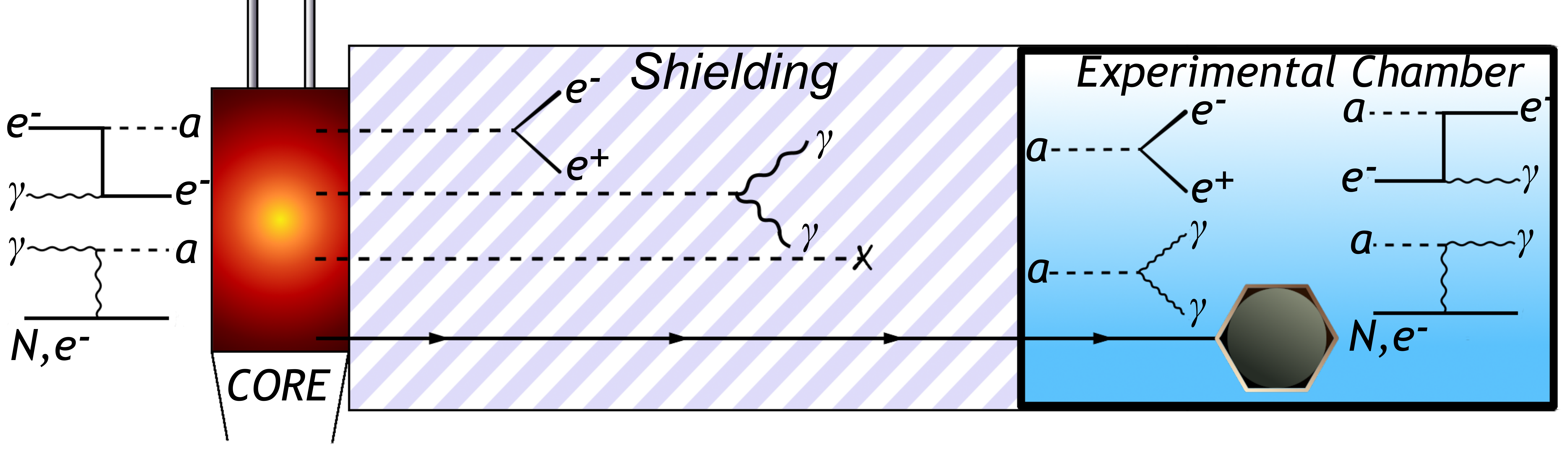}
    \caption{Cartoon of the ALPs and their production (left), scattering, and decay possibilities (right) at a reactor neutrino experiment. The ALP may decay inside the shielding and evade detection (dashed lines). ALPs that free stream through the shielding (solid line) may be detected via the inverse Primakoff and Compton scattering channels and decay channels. The detector would be housed inside a hermetic shielding to further reduce the gamma and neutron backgrounds (solid blue).}
    \label{fig:feynman}
\end{figure*}

Due to the photon coupling, ALPs can be produced through the Primakoff process $\gamma(p_1) + A(p_2) \rightarrow a(k_1) + A(k_2)$ \cite{Pirmakoff:1951pj}, where $A$ is an atomic target (Fig. \ref{fig:feynman}). This interaction proceeds through a $t$-channel photon exchange whose rate is governed by the strength of the coupling $g_{a\gamma\gamma}$. This process is enhanced by the coherency factor $Z^2$ where $Z$ is the atomic number. The forward scattering differential cross-section is  \cite{Tsai:1986tx,Aloni:2019ruo}
\bea
\label{eq:Primakoff}
\frac{d\sigma_P^p}{d\cos\theta} = \frac{1}{4}g_{a\gamma\gamma}^2\alpha Z^2F^2(t)\frac{|\vec{p}_a|^4\sin^2\theta}{t^2}
\eea
We will use superscripts $p$ and $d$ to distinguish between production and detection cross-sections, respectively. Here $\alpha = e^2/(4\pi)$ is the standard electromagnetic fine structure constant, $F^2(t)$ contains the atomic and nuclear form factors, and $|\vec{p}_a|$ is the magnitude of the outgoing three-momentum of the ALP at the angle $\theta$ relative to the incident photon momentum. The square of the four-momentum transfer is given by $t = (p_1-k_1)^2=m_a^2 + E_\gamma(E_a -|\vec{p}_a|\cos\theta)$ for a photon of incident energy $E_\gamma$ that produces an ALP of energy $E_a$ and mass $m_a$.

ALPs can also be produced through an $s$- plus $u$-channel Compton-like scattering process on electron targets $\gamma + e^- \rightarrow a + e^-$ which has a differential cross-section \cite{Brodsky:1986mi,Chakrabarty:2019kdd,Gondolo:2008dd}
\bea
\label{eq:ae_production}
\frac{d\sigma_C^p}{dx} = \frac{Z \pi g_{aee}^2 \alpha x}{4\pi(s- m_e^2)(1-x)}&& \hspace{-0.5cm}\bigg[x -\, \, \frac{2m_a^2 s}{(s-m_e^2)^2} \\
&+& \frac{2m_a^2}{(s-m_e^2)^2} \bigg(\frac{m_e^2}{1-x} +\frac{m_a^2}{x}\bigg)\bigg] \nonumber
\eea
where $m_e$ is the electron mass, $s$ is the usual Mandelstam variable ($s-m_e^2 = 2E_\gamma m_e$ in the electron rest frame), and $x$ is the fractional light cone momentum, which can take values between 0 and 1.
In the laboratory frame, one may perform a change of variables using $x=1-\dfrac{E_a}{E_\gamma}+\dfrac{m_a^2}{2E_\gamma m_e}$.

Within the framework adopted here, once produced, the ALP can generate a detectable signal in several ways. The ALP could decay to two photons or an electron-positron pair with the well-known decay widths
\begin{eqnarray}
\label{eq:decay}
\Gamma (a \to \gamma \gamma) &=& \dfrac{g_{a\gamma\gamma}^2 m_a^3}{64\pi}\,, \\
\label{eq:decay_ee}
\Gamma (a \to e^+ e^-) &=& \dfrac{g_{aee}^2 m_a}{8\pi} \sqrt{1 - \dfrac{4 m_e^2}{m_a^2}}
\end{eqnarray}
which, in conjunction with the ALP kinetic energy, fix the decay length. Secondly, the ALP could be detected through the inverse Primakoff process $a + A \rightarrow \gamma + A$, which has the same differential cross-section as in Eq.~\eqref{eq:Primakoff}, with the alteration that the front-factor 1/4 becomes 1/2 due to the initial spin states including a spin-0 ALP rather than a spin-1 photon. Therefore, for non-zero $g_{a\gamma\gamma}$, the production (via Primakoff) and the scattering (inverse) cross-sections involving both electron and nucleus  in the atom  have a $Z^2$ enhancement~\cite{Tsai:1986tx}. Finally, the ALP could interact with electrons through the inverse Compton-like process, $a+e^- \rightarrow \gamma + e^-$, which produces photons from electron bremsstrahlung as well as electron recoils for non-zero $g_{aee}$ where the enhancement factor is $Z$ as in the production case shown in the Eq.~\eqref{eq:ae_production}. This process has a differential cross-section of the form \cite{Avignone:1988bv,Bellini:2008zza}
\begin{align}
\label{eq:ae_scattering}
\frac{d\sigma_C^d}{d\Omega} = \frac{Z g_{aee}^2\alpha E_\gamma}{8\pi m_e^2 \mid \vec{p}_a \mid}\bigg(1 & + \frac{4m_e^2 E^2_\gamma}{(2m_e E_a + m_a^2)^2} - \frac{4m_e E_\gamma}{(2m_e E_a + m_a^2)}
\nonumber \\
&-\frac{4m_a^2 \mid \vec{p}_a \mid^2m_e E_\gamma\sin^2\theta}{(2m_e E_a + m_a^2)^3}\bigg)
\end{align}\\

\noindent {\bf \emph{The} \emph{Experimental Setup. - }}Let us first discuss MINER as a baseline example to consider for an ALP search. The MINER experiment consists of SuperCDMS-style cryogenic germanium detectors situated at 4.5 m from the core of a TRIGA type 1-MW reactor with low enriched ${}^{235}$U at the NSC (the reactor-detector system allows for closer proximity down to $\sim 2$ m for the next phase). Though the experiment was established for detection of CE$\nu$NS, it is also ideally situated for ALP searches in previously unexplored regions of $m_a-g_{a\gamma\gamma}$
parameter space. This is due to the combination of a substantial photon flux of $10^{19}~\gamma/$s from the reactor, the nearness of the detectors, their low-threshold sensitivity, and detection via both scattering and decay channels. As an example of the reach for this experimental layout, for $m_a = 1$ MeV and $g_{a\gamma\gamma} = 10^{-6}$ MeV$^{-1}$,
the photon flux from ALP decay will be approximately $13.6~\rm{cm}^{-2}\rm{s}^{-1}$, with an ALP flux of $72.0~\rm{cm}^{-2}\rm{s}^{-1}$. Depending on the choice of $m_a$ and $g_{a\gamma\gamma}$, 
the photon rates may vary in comparison between ALP scattering and decay.

Estimation of the ALP signal rate is performed as follows. We take a reactor photon flux from MINER, described in Ref.~\cite{Agnolet:2016zir}, which we restrict to $>25$ keV in energy due to the binning of the background simulation, taken at the reactor core. We expect the integrated reactor flux to scale linearly with thermal power of the core, which provides rudimentary means of extrapolating this flux to the other GW-scale reactors in our consideration. We then convolve the flux with the Primakoff or Compton cross-sections to produce ALPs from photons scattering with the core material, in this case approximated by a core of pure thorium ($Z=90$, averaging across atomic numbers in the core). ALPs are then allowed to propagate through the shielding material until they either decay in flight or scatter off the detector material. 

The convolution performed here is similar to the one in the TEXONO analysis, except in this case the production mechanism via Primakoff or Compton conversion imposes a branching ratio $\Gamma_a / \Gamma \equiv \sigma_a^p/(\sigma_a^p + \sigma_{SM})$. Here $\sigma_a^p = \sigma_C^p, \sigma_P^p$ is the total Compton or Primakoff axion-production cross-section, respectively, and $\sigma_{SM}$ represents the total photon scattering cross-section against core material taken from the Photon Cross Sections Database \cite{NIST:2019}. The event yield $S$ from ALP scattering is therefore given by, in the Primakoff case,
\bea
S = N_T \int \sigma_P^d \cdot \dfrac{\Gamma_a}{\Gamma} \cdot P_{surv} \cdot \dfrac{d\Phi_\gamma}{dE_\gamma}\,\, dE_\gamma
\eea
where $N_T$ is the number of target atoms, $\sigma_P^d$ is the Primakoff scattering cross-section in the detector, $\frac{d\Phi_\gamma}{dE_\gamma}$ is the differential photon flux at the detector, and $P_{surv} = \exp[-\ell_d m_a / (p_a \tau)]$ is the axion survival probability for a core-detector proximity $\ell_d$, axion lifetime $\tau$, and momentum $p_a$.
In the Compton case, we must also include the differential probability of producing an axion with energy $E_a$, giving an additional factor $\frac{1}{\sigma_C}\frac{d\sigma_C}{dE_a}$ and an additional integration over axion energies (unlike the Primakoff case, in which the photon energy is coherently converted into axion production, where we have which we have implicitly integrated out the factor of $\delta(E_a - E_\gamma)$ in Eq.~\eqref{eq:Primakoff}). To keep the analysis simple, no ALP flux attenuation is applied from scattering inside the shielding; however, we also assume no ALP production inside the shielding, nor do we include other channels of ALP production (e.g., axion bremsstrahlung) inside the core, leaving the signal yield estimate on the conservative side.

Diphoton and electron-positron pair production from ALP decay may also contribute to the event yield, but the ALP must bypass the shielding sections in order for the decay products to be seen by the detector; hence, after taking into account the number of axions that survive the shielding sections, we apply the additional decay probability
$P_{decay} = 1 - \exp\left[-\Delta\ell m_a/(p_a \tau)\right]$
for a fiducial detector length $\Delta\ell$.

The estimated detector and reactor specifications relevant to the signal estimation (thermal power, core proximities, background rates and exposures) for MINER, $\nu$-cleus, CONNIE and CONUS are listed in Table~\ref{tab:benchmarks}. Since these experiments are specific to neutrino scattering, approximate background rates are quoted in their appropriate regions of interest (ROI), determined by the endpoints of the neutrino-nucleus recoil response in the range of 1-4 keVnr, depending on the detector material. Although this energy range is small compared to the signal region for an ALP search that one may consider, the rates should give a good approximation of the total background, which substantially attenuates beyond these nuclear endpoints. Then, by the 2.6 MeV thallium endpoint, all radiochemical backgrounds should be absent. The sensitivities to $g_{a\gamma\gamma}$ and $g_{aee}$ should be stable to uncertainties on these backgrounds up to an order-of-magnitude, since the signal yield is proportional to four powers of the coupling. Additionally, backgrounds arising from radiochemical sources external to the detector as well as reactor photons that pass through the shielding may be further suppressed using a scintillating module that encases the detector within the experimental chamber. Photons that pass through this layer may then be vetoed in the event trigger in favor of ALP-like signals in which invisibly enter the detector volume. This is a possibility at the MINER experiment, but in this work we place all experiments on the same footing and assume no photon veto.

\begin{figure}
    \centering
    \includegraphics[width=8.4cm]{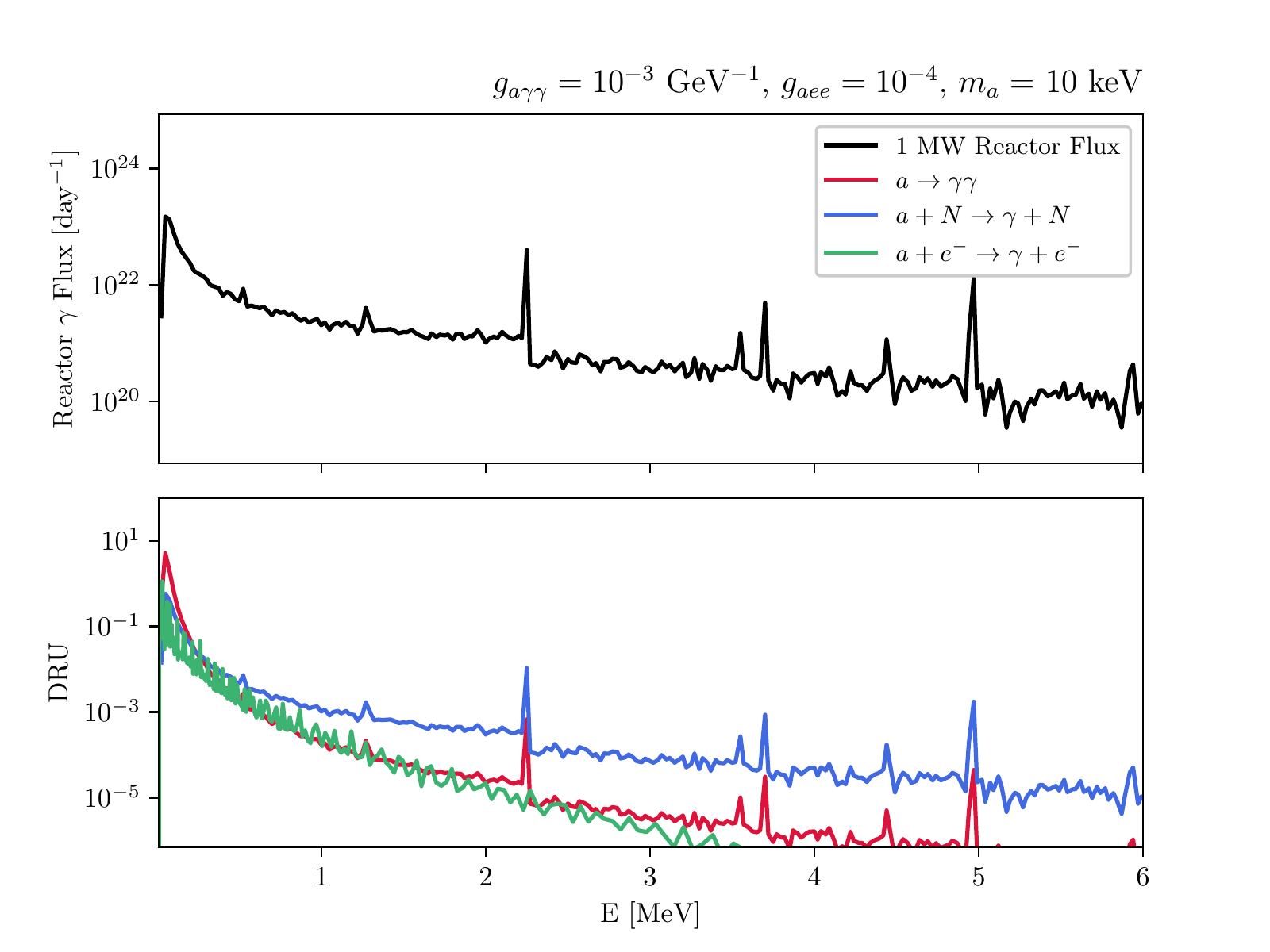}
    \caption{The 1-MW MINER photon flux at the reactor core is shown in black (top). The subsequent scatter and decay rates from ALPs in the detector volume are shown (bottom) for individually chosen values of $g_{a\gamma\gamma}$ and $g_{aee}$ for a 10-keV ALP. Perfect detector efficiency and energy reconstruction is assumed. Expected backgrounds are omitted, as they require dedicated analyses specialized to each detector.}
    \label{fig:miner_ge_rates}
\end{figure}

The detected photon spectrum from ALPs, produced and detected in Primakoff and Compton channels, or decaying to diphoton pairs in the detector, is shown in Fig.~\ref{fig:miner_ge_rates} for a 1-MW reactor. As from Eq.~\eqref{eq:Primakoff}, the scattering spectrum pictured is independent of $m_a$ in the forward limit, while the ALP decay-driven photon rate is dependent on both $m_a$ and $g_{a\gamma\gamma}$ via Eq.~\eqref{eq:decay}. The expected backgrounds are not pictured, but they are expected to attenuate quickly by the $2.6$-MeV thallium endpoint. Since we consider a variety of experiments in this analysis, we will avoid assumptions about the background shape and conservatively opt for a single-binned treatment of the signal and background.

\noindent {\bf \emph{Results. - }}Having set the stage for the ALP search at an array of reactor neutrino experiments, we are now in a position to present its reach on both ALP-photon and ALP-electron couplings over a range of ALP masses. We evaluate limits on the ALP signal sensitivity for MINER, CONNIE, CONUS and $\nu$-cleus experiments keeping only one coupling non-zero at a time.

\begin{table}[h]
    \centering
    \caption{Approximate specifications for the reactor and detector benchmarks are summarized from Refs.~\cite{AristizabalSierra:2019hcm,Aguilar-Arevalo:2016khx,Agnolet:2016zir,Strauss:2017cuu,Buck:2020opf}. Background rates in DRU (kg$^{-1}$keV$^{-1}$ day$^{-1}$) are listed, and are based on the rates that appear in the ROI of each respective experiment. Exposures are based on a 3-year run period.}
    \begin{tabular}{|l|c|c|c|c|}
         \hline
         Experiment & \thead{Core\\Thermal\\Power} & \thead{Core\\Proximity\\(m)} & \thead{Bkg Rate\\in ROI\\(DRU)} & \thead{Exposure\\(kg$\cdot$days)} \\
         \hline
         MINER (Ge) & 1 MW & 2.25 & 100 & 4000 \\
        $\nu$-cleus (CaWO$_4$) & 4 GW & 40 & 100 & 10 \\
         CONNIE (Si CCD) & 4 GW & 30 & 700 & 100 \\
         CONUS (Ge PPC) & 4 GW & 17 & 100 & 4000 \\
         \hline
    \end{tabular}
    \label{tab:benchmarks}
\end{table}

As a conservative evaluation of the sensitivities for each benchmark experiment, we calculate the projected limits on the ALP mass and couplings via a single energy bin analysis using $\kappa = \dfrac{N_s}{\sqrt{N_s + N_b}}$ as a test statistic where $N_s$ and $N_b$ are the integrated signal and background events, respectively. Given more precise knowledge of the expected background spectra, reactor photon flux, and shielding geometry, an energy-binned analysis could improve the sensitivity to ALP signals by taking advantage of the vanishing background at 2.6 MeV.

\begin{figure}
    \centering
    \includegraphics[width=8.4cm]{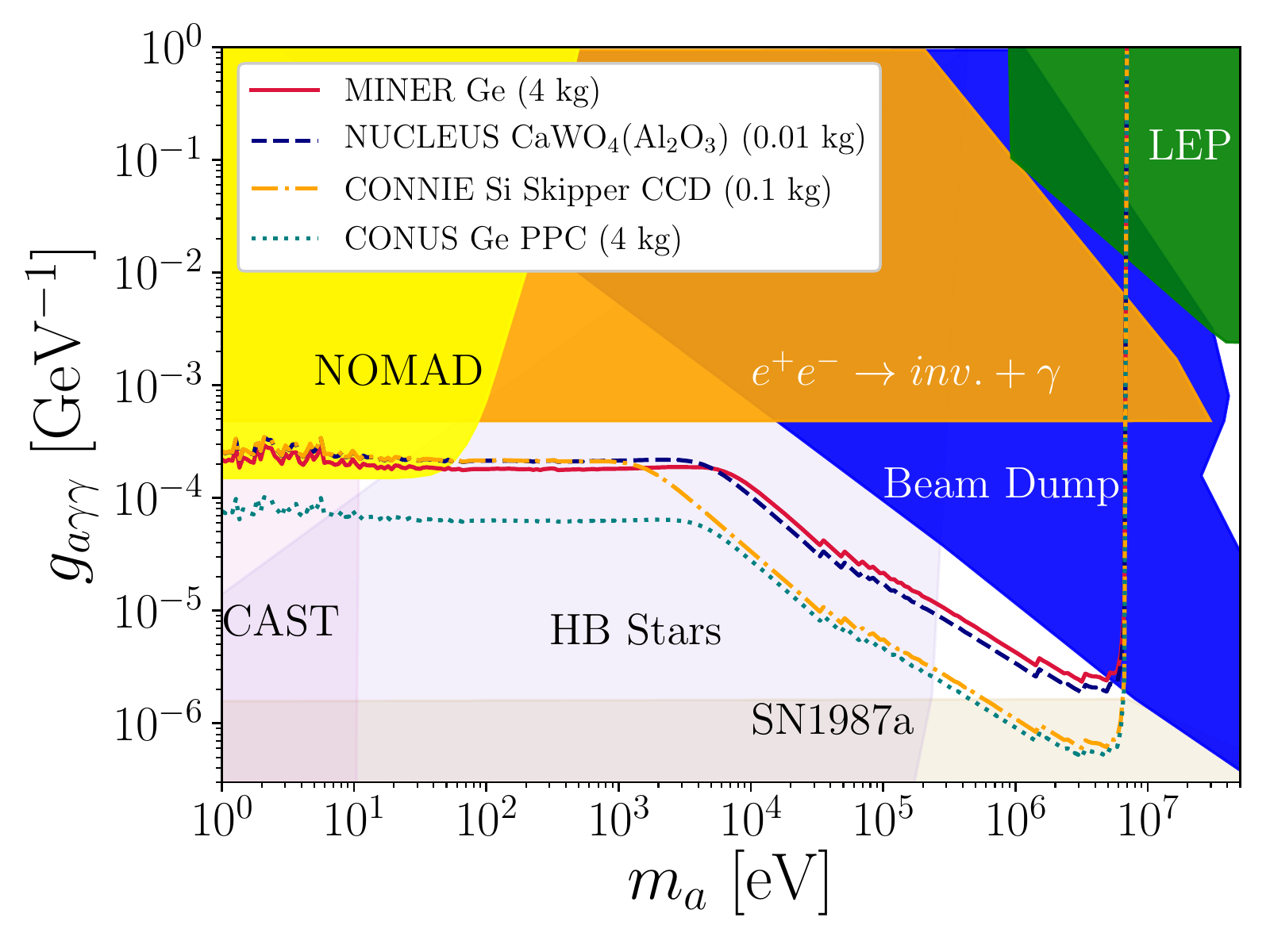}
    \includegraphics[width=8.4cm]{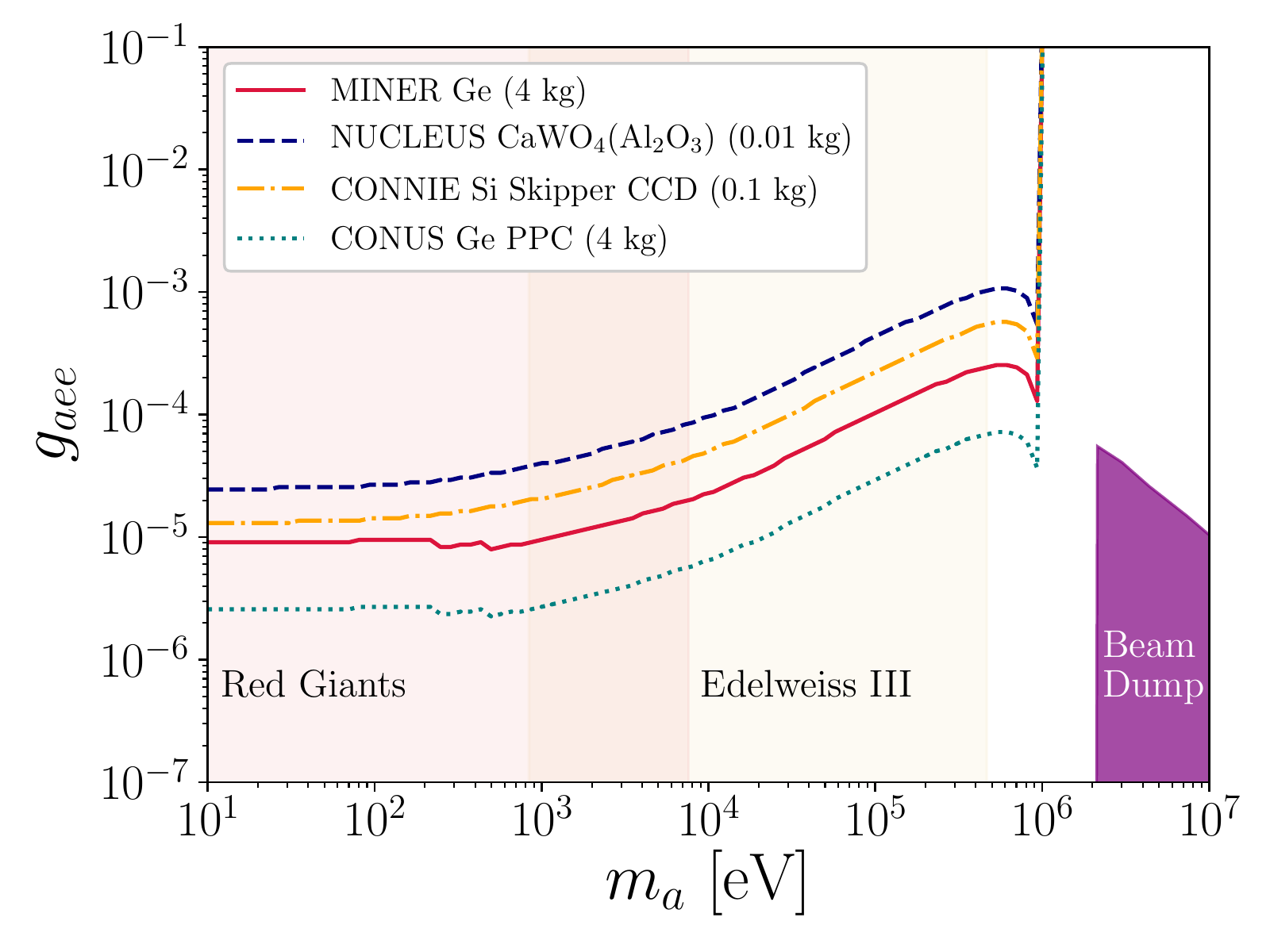}
    \caption{Limits for $\kappa=2$ ($\simeq$ 95\% C.L.), after a 3-year exposure, are derived on the ALP-photon (top) and ALP-electron (bottom) couplings $g_{a\gamma\gamma}$ and $g_{aee}$ as a function of ALP mass $m_a$ for the MINER, $\nu$-cleus, CONNIE, and CONUS benchmarks. Astrophysically-derived constraints are shown with light shading. The limit on each coupling assumes all other ALP couplings to be negligible.}
    \label{fig:miner_limits}
\end{figure}

Fig.~\ref{fig:miner_limits} (top) shows the resulting $\kappa = 2$ contours ($\simeq 95\%$ C.L.) on $(m_a, g_{a\gamma\gamma})$ for MINER, CONUS, CONNIE and $\nu$-cleus for ALPs coupled purely to photons. The flat limit for $m_a \leq 10^4$ eV is set by ALP scattering in the detector material and is $m_a$-independent (the reactor photon flux producing the axion is also constant in this energy range, originating from the single energy bin below 25 keV), while the limit peaked at $m_a \simeq 4$ MeV is set by the $a \to \gamma \gamma$ rate which depends on the distance from the flux source and the ALP decay length. The sharp loss in sensitivity by $m_a \simeq$ 10 MeV is due to the high-energy endpoint of the reactor photon flux sample that was used, which is quickly falling to zero by $E_\gamma \simeq 10$ MeV, and therefore places an upper limit on the producible axion mass. MINER also plans to use a 200-kg CsI detector in the upcoming future. However, since  the background for a CsI detector is not given for a reactor experiment, we are not providing any projection for this future plan.

Fig.~\ref{fig:miner_limits} (bottom) shows similar limits on $(m_a, g_{aee})$ for MINER, CONUS, CONNIE and $\nu$-cleus for ALPs coupled purely to electrons. The limits are dominated by ALPs scattering off the detector material. Contributions from ALP decay $a \to e^+ e^-$ require $m_a > 2 m_e \simeq 1$~MeV, but $m_a$ cannot be too heavy in order to satisfy the production threshold $s \geq (m_a + m_e)^2$. In this narrow range of $m_a$, $g_{aee}$ is either too large and $a$ decays before reaching the detector, or $g_{aee}$ is too small to give a statistically significant scattering yield. In the low $m_a$ limit, sensitivity flattens out as the ALP-electron scattering differential cross-section (Eq.~\eqref{eq:ae_scattering}) becomes $m_a$-independent (again, the reactor photon flux is constant arising from the single energy bin below 25~keV).

Among the reactor-based experiments, CONUS has the best sensitivity as it has the closest core-detector proximity and largest exposure of the GW-scale reactor experiments. CONNIE and $\nu$-cleus sensitivities are smaller than MINER and CONUS due to their smaller detector masses and relatively larger separations from the photon source at the core. We expect all of these experiments can beat the existing best constraints from the laboratory all the way down to $m_a = 100$~eV for photon couplings. From $m_a = 1$~MeV to the massless limit, our projections exceed all laboratory constraints on axion-electron couplings. Lastly, while constraints from astrophysics and cosmology may be evaded by a wide variety of models as discussed previously, we also show relevant existing astrophysical constraints~\cite{Armengaud:2018cuy,Bauer:2018uxu}, colored with light shading, in Fig.~\ref{fig:miner_limits}. Even for ALP models where those constraints apply, we project coverage over new parameter space, in particular over the ``cosmological triangle" of unexplored parameter space formed between the beam dump, HB stars, and SN1987a exclusions.

This work has demonstrated the exciting and new possibility of highly sensitive ALP searches from ALP scattering processes with currently existing low-threshold detectors at MW nuclear reactor facilities such as MINER, as well as GW reactor experiments, e.g., CONUS, CONNIE, and $\nu$-cleus.
In addition to reactor-based searches, stopped-pion experiments also have a high photon flux which can be leveraged for similar ALP searches \footnote{Work in progress.}. 

\medskip

\noindent {\bf \emph{Acknowledgement. - }} J.B.D. acknowledges support from the National Science Foundation under Grant No. NSF PHY182080, useful discussions with J. Newstead, and S. Sabharwal for a calculational check. B.D., D.K., and S.L. acknowledge support from DOE Grant DE-SC0010813. K.S. acknowledges support from DOE Grant DE-SC0009956. R.M. acknowledges support from DOE grants DE-SC0018981 and DE-SC0017859. S.L. and A.T. thank the Mitchell Institute for Fundamental Physics and Astronomy for support. We thank A. Kubik, S. Rajendran,  K. Scholberg and L. Strigari for useful discussions.

\bibliography{ALP}

\begin{thebibliography}{69}%
\makeatletter
\providecommand \@ifxundefined [1]{%
 \@ifx{#1\undefined}
}%
\providecommand \@ifnum [1]{%
 \ifnum #1\expandafter \@firstoftwo
 \else \expandafter \@secondoftwo
 \fi
}%
\providecommand \@ifx [1]{%
 \ifx #1\expandafter \@firstoftwo
 \else \expandafter \@secondoftwo
 \fi
}%
\providecommand \natexlab [1]{#1}%
\providecommand \enquote  [1]{``#1''}%
\providecommand \bibnamefont  [1]{#1}%
\providecommand \bibfnamefont [1]{#1}%
\providecommand \citenamefont [1]{#1}%
\providecommand \href@noop [0]{\@secondoftwo}%
\providecommand \href [0]{\begingroup \@sanitize@url \@href}%
\providecommand \@href[1]{\@@startlink{#1}\@@href}%
\providecommand \@@href[1]{\endgroup#1\@@endlink}%
\providecommand \@sanitize@url [0]{\catcode `\\12\catcode `\$12\catcode
  `\&12\catcode `\#12\catcode `\^12\catcode `\_12\catcode `\%12\relax}%
\providecommand \@@startlink[1]{}%
\providecommand \@@endlink[0]{}%
\providecommand \url  [0]{\begingroup\@sanitize@url \@url }%
\providecommand \@url [1]{\endgroup\@href {#1}{\urlprefix }}%
\providecommand \urlprefix  [0]{URL }%
\providecommand \Eprint [0]{\href }%
\providecommand \doibase [0]{http://dx.doi.org/}%
\providecommand \selectlanguage [0]{\@gobble}%
\providecommand \bibinfo  [0]{\@secondoftwo}%
\providecommand \bibfield  [0]{\@secondoftwo}%
\providecommand \translation [1]{[#1]}%
\providecommand \BibitemOpen [0]{}%
\providecommand \bibitemStop [0]{}%
\providecommand \bibitemNoStop [0]{.\EOS\space}%
\providecommand \EOS [0]{\spacefactor3000\relax}%
\providecommand \BibitemShut  [1]{\csname bibitem#1\endcsname}%
\let\auto@bib@innerbib\@empty
\bibitem [{\citenamefont {Peccei}\ and\ \citenamefont
  {Quinn}(1977)}]{Peccei:1977hh}%
  \BibitemOpen
  \bibfield  {author} {\bibinfo {author} {\bibfnamefont {R.~D.}\ \bibnamefont
  {Peccei}}\ and\ \bibinfo {author} {\bibfnamefont {H.~R.}\ \bibnamefont
  {Quinn}},\ }\href {\doibase 10.1103/PhysRevLett.38.1440} {\bibfield
  {journal} {\bibinfo  {journal} {Phys. Rev. Lett.}\ }\textbf {\bibinfo
  {volume} {38}},\ \bibinfo {pages} {1440} (\bibinfo {year}
  {1977})}\BibitemShut {NoStop}%
\bibitem [{\citenamefont {Wilczek}(1978)}]{Wilczek:1977pj}%
  \BibitemOpen
  \bibfield  {author} {\bibinfo {author} {\bibfnamefont {F.}~\bibnamefont
  {Wilczek}},\ }\href {\doibase 10.1103/PhysRevLett.40.279} {\bibfield
  {journal} {\bibinfo  {journal} {Phys. Rev. Lett.}\ }\textbf {\bibinfo
  {volume} {40}},\ \bibinfo {pages} {279} (\bibinfo {year} {1978})}\BibitemShut
  {NoStop}%
\bibitem [{\citenamefont {Weinberg}(1978)}]{Weinberg:1977ma}%
  \BibitemOpen
  \bibfield  {author} {\bibinfo {author} {\bibfnamefont {S.}~\bibnamefont
  {Weinberg}},\ }\href {\doibase 10.1103/PhysRevLett.40.223} {\bibfield
  {journal} {\bibinfo  {journal} {Phys. Rev. Lett.}\ }\textbf {\bibinfo
  {volume} {40}},\ \bibinfo {pages} {223} (\bibinfo {year} {1978})}\BibitemShut
  {NoStop}%
\bibitem [{\citenamefont {Duffy}\ and\ \citenamefont {van
  Bibber}(2009)}]{Duffy:2009ig}%
  \BibitemOpen
  \bibfield  {author} {\bibinfo {author} {\bibfnamefont {L.~D.}\ \bibnamefont
  {Duffy}}\ and\ \bibinfo {author} {\bibfnamefont {K.}~\bibnamefont {van
  Bibber}},\ }\href {\doibase 10.1088/1367-2630/11/10/105008} {\bibfield
  {journal} {\bibinfo  {journal} {New J. Phys.}\ }\textbf {\bibinfo {volume}
  {11}},\ \bibinfo {pages} {105008} (\bibinfo {year} {2009})},\ \Eprint
  {http://arxiv.org/abs/0904.3346} {arXiv:0904.3346 [hep-ph]} \BibitemShut
  {NoStop}%
\bibitem [{\citenamefont {Marsh}(2016)}]{Marsh:2015xka}%
  \BibitemOpen
  \bibfield  {author} {\bibinfo {author} {\bibfnamefont {D.~J.~E.}\
  \bibnamefont {Marsh}},\ }\href {\doibase 10.1016/j.physrep.2016.06.005}
  {\bibfield  {journal} {\bibinfo  {journal} {Phys. Rept.}\ }\textbf {\bibinfo
  {volume} {643}},\ \bibinfo {pages} {1} (\bibinfo {year} {2016})},\ \Eprint
  {http://arxiv.org/abs/1510.07633} {arXiv:1510.07633 [astro-ph.CO]}
  \BibitemShut {NoStop}%
\bibitem [{\citenamefont {Battaglieri}\ \emph {et~al.}(2017)\citenamefont
  {Battaglieri} \emph {et~al.}}]{Battaglieri:2017aum}%
  \BibitemOpen
  \bibfield  {author} {\bibinfo {author} {\bibfnamefont {M.}~\bibnamefont
  {Battaglieri}} \emph {et~al.},\ }in\ \href
  {http://lss.fnal.gov/archive/2017/conf/fermilab-conf-17-282-ae-ppd-t.pdf}
  {\emph {\bibinfo {booktitle} {{U.S. Cosmic Visions: New Ideas in Dark Matter
  College Park, MD, USA, March 23-25, 2017}}}}\ (\bibinfo {year} {2017})\
  \Eprint {http://arxiv.org/abs/1707.04591} {arXiv:1707.04591 [hep-ph]}
  \BibitemShut {NoStop}%
\bibitem [{\citenamefont {Zioutas}\ \emph {et~al.}(1999)\citenamefont {Zioutas}
  \emph {et~al.}}]{Zioutas:1998cc}%
  \BibitemOpen
  \bibfield  {author} {\bibinfo {author} {\bibfnamefont {K.}~\bibnamefont
  {Zioutas}} \emph {et~al.},\ }\href {\doibase 10.1016/S0168-9002(98)01442-9}
  {\bibfield  {journal} {\bibinfo  {journal} {Nucl. Instrum. Meth.}\ }\textbf
  {\bibinfo {volume} {A425}},\ \bibinfo {pages} {480} (\bibinfo {year}
  {1999})},\ \Eprint {http://arxiv.org/abs/astro-ph/9801176}
  {arXiv:astro-ph/9801176 [astro-ph]} \BibitemShut {NoStop}%
\bibitem [{\citenamefont {Anastassopoulos}\ \emph {et~al.}(2017)\citenamefont
  {Anastassopoulos} \emph {et~al.}}]{Anastassopoulos:2017ftl}%
  \BibitemOpen
  \bibfield  {author} {\bibinfo {author} {\bibfnamefont {V.}~\bibnamefont
  {Anastassopoulos}} \emph {et~al.} (\bibinfo {collaboration} {CAST}),\ }\href
  {\doibase 10.1038/nphys4109} {\bibfield  {journal} {\bibinfo  {journal}
  {Nature Phys.}\ }\textbf {\bibinfo {volume} {13}},\ \bibinfo {pages} {584}
  (\bibinfo {year} {2017})},\ \Eprint {http://arxiv.org/abs/1705.02290}
  {arXiv:1705.02290 [hep-ex]} \BibitemShut {NoStop}%
\bibitem [{\citenamefont {Irastorza}\ \emph {et~al.}(2013)\citenamefont
  {Irastorza} \emph {et~al.}}]{Irastorza:2013dav}%
  \BibitemOpen
  \bibfield  {author} {\bibinfo {author} {\bibfnamefont {I.}~\bibnamefont
  {Irastorza}} \emph {et~al.} (\bibinfo {collaboration} {IAXO}),\ }\href@noop
  {} {\  (\bibinfo {year} {2013})}\BibitemShut {NoStop}%
\bibitem [{\citenamefont {Kahn}\ \emph {et~al.}(2016)\citenamefont {Kahn},
  \citenamefont {Safdi},\ and\ \citenamefont {Thaler}}]{Kahn:2016aff}%
  \BibitemOpen
  \bibfield  {author} {\bibinfo {author} {\bibfnamefont {Y.}~\bibnamefont
  {Kahn}}, \bibinfo {author} {\bibfnamefont {B.~R.}\ \bibnamefont {Safdi}}, \
  and\ \bibinfo {author} {\bibfnamefont {J.}~\bibnamefont {Thaler}},\ }\href
  {\doibase 10.1103/PhysRevLett.117.141801} {\bibfield  {journal} {\bibinfo
  {journal} {Phys. Rev. Lett.}\ }\textbf {\bibinfo {volume} {117}},\ \bibinfo
  {pages} {141801} (\bibinfo {year} {2016})},\ \Eprint
  {http://arxiv.org/abs/1602.01086} {arXiv:1602.01086 [hep-ph]} \BibitemShut
  {NoStop}%
\bibitem [{\citenamefont {Salemi}(2019)}]{Salemi:2019xgl}%
  \BibitemOpen
  \bibfield  {author} {\bibinfo {author} {\bibfnamefont {C.~P.}\ \bibnamefont
  {Salemi}} (\bibinfo {collaboration} {ABRACADABRA}),\ }in\ \href@noop {}
  {\emph {\bibinfo {booktitle} {{54th Rencontres de Moriond on Electroweak
  Interactions and Unified Theories (Moriond EW 2019) La Thuile, Italy, March
  16-23, 2019}}}}\ (\bibinfo {year} {2019})\ \Eprint
  {http://arxiv.org/abs/1905.06882} {arXiv:1905.06882 [hep-ex]} \BibitemShut
  {NoStop}%
\bibitem [{\citenamefont {Asztalos}\ \emph {et~al.}(2001)\citenamefont
  {Asztalos} \emph {et~al.}}]{Asztalos:2001tf}%
  \BibitemOpen
  \bibfield  {author} {\bibinfo {author} {\bibfnamefont {S.~J.}\ \bibnamefont
  {Asztalos}} \emph {et~al.} (\bibinfo {collaboration} {ADMX}),\ }\href
  {\doibase 10.1103/PhysRevD.64.092003} {\bibfield  {journal} {\bibinfo
  {journal} {Phys. Rev. D}\ }\textbf {\bibinfo {volume} {64}},\ \bibinfo
  {pages} {092003} (\bibinfo {year} {2001})}\BibitemShut {NoStop}%
\bibitem [{\citenamefont {Du}\ \emph {et~al.}(2018)\citenamefont {Du} \emph
  {et~al.}}]{Du:2018uak}%
  \BibitemOpen
  \bibfield  {author} {\bibinfo {author} {\bibfnamefont {N.}~\bibnamefont {Du}}
  \emph {et~al.} (\bibinfo {collaboration} {ADMX}),\ }\href {\doibase
  10.1103/PhysRevLett.120.151301} {\bibfield  {journal} {\bibinfo  {journal}
  {Phys. Rev. Lett.}\ }\textbf {\bibinfo {volume} {120}},\ \bibinfo {pages}
  {151301} (\bibinfo {year} {2018})},\ \Eprint
  {http://arxiv.org/abs/1804.05750} {arXiv:1804.05750 [hep-ex]} \BibitemShut
  {NoStop}%
\bibitem [{\citenamefont {Jackson~Kimball}\ \emph {et~al.}(2017)\citenamefont
  {Jackson~Kimball} \emph {et~al.}}]{JacksonKimball:2017elr}%
  \BibitemOpen
  \bibfield  {author} {\bibinfo {author} {\bibfnamefont {D.~F.}\ \bibnamefont
  {Jackson~Kimball}} \emph {et~al.},\ }\href@noop {} {\  (\bibinfo {year}
  {2017})},\ \Eprint {http://arxiv.org/abs/1711.08999} {arXiv:1711.08999
  [physics.ins-det]} \BibitemShut {NoStop}%
\bibitem [{\citenamefont {Brubaker}\ \emph {et~al.}(2017)\citenamefont
  {Brubaker} \emph {et~al.}}]{Brubaker:2016ktl}%
  \BibitemOpen
  \bibfield  {author} {\bibinfo {author} {\bibfnamefont {B.~M.}\ \bibnamefont
  {Brubaker}} \emph {et~al.},\ }\href {\doibase 10.1103/PhysRevLett.118.061302}
  {\bibfield  {journal} {\bibinfo  {journal} {Phys. Rev. Lett.}\ }\textbf
  {\bibinfo {volume} {118}},\ \bibinfo {pages} {061302} (\bibinfo {year}
  {2017})},\ \Eprint {http://arxiv.org/abs/1610.02580} {arXiv:1610.02580
  [astro-ph.CO]} \BibitemShut {NoStop}%
\bibitem [{\citenamefont {Droster}\ and\ \citenamefont {van
  Bibber}(2019)}]{Droster:2019fur}%
  \BibitemOpen
  \bibfield  {author} {\bibinfo {author} {\bibfnamefont {A.}~\bibnamefont
  {Droster}}\ and\ \bibinfo {author} {\bibfnamefont {K.}~\bibnamefont {van
  Bibber}} (\bibinfo {collaboration} {HAYSTAC}),\ }in\ \href@noop {} {\emph
  {\bibinfo {booktitle} {{13th Conference on the Intersections of Particle and
  Nuclear Physics (CIPANP 2018) Palm Springs, California, USA, May 29-June 3,
  2018}}}}\ (\bibinfo {year} {2019})\ \Eprint {http://arxiv.org/abs/1901.01668}
  {arXiv:1901.01668 [physics.ins-det]} \BibitemShut {NoStop}%
\bibitem [{\citenamefont {Spector}(2019)}]{Spector:2019ooq}%
  \BibitemOpen
  \bibfield  {author} {\bibinfo {author} {\bibfnamefont {A.}~\bibnamefont
  {Spector}} (\bibinfo {collaboration} {ALPS}),\ }in\ \href@noop {} {\emph
  {\bibinfo {booktitle} {{14th Patras Workshop on Axions, WIMPs and WISPs
  (AXION-WIMP 2018) (PATRAS 2018) Hamburg, Germany, June 18-22, 2018}}}}\
  (\bibinfo {year} {2019})\ \Eprint {http://arxiv.org/abs/1906.09011}
  {arXiv:1906.09011 [physics.ins-det]} \BibitemShut {NoStop}%
\bibitem [{\citenamefont {Melissinos}(2009)}]{Melissinos:2008vn}%
  \BibitemOpen
  \bibfield  {author} {\bibinfo {author} {\bibfnamefont {A.~C.}\ \bibnamefont
  {Melissinos}},\ }\href {\doibase 10.1103/PhysRevLett.102.202001} {\bibfield
  {journal} {\bibinfo  {journal} {Phys. Rev. Lett.}\ }\textbf {\bibinfo
  {volume} {102}},\ \bibinfo {pages} {202001} (\bibinfo {year} {2009})},\
  \Eprint {http://arxiv.org/abs/0807.1092} {arXiv:0807.1092 [hep-ph]}
  \BibitemShut {NoStop}%
\bibitem [{\citenamefont {DeRocco}\ and\ \citenamefont
  {Hook}(2018)}]{DeRocco:2018jwe}%
  \BibitemOpen
  \bibfield  {author} {\bibinfo {author} {\bibfnamefont {W.}~\bibnamefont
  {DeRocco}}\ and\ \bibinfo {author} {\bibfnamefont {A.}~\bibnamefont {Hook}},\
  }\href {\doibase 10.1103/PhysRevD.98.035021} {\bibfield  {journal} {\bibinfo
  {journal} {Phys. Rev. D}\ }\textbf {\bibinfo {volume} {98}},\ \bibinfo
  {pages} {035021} (\bibinfo {year} {2018})},\ \Eprint
  {http://arxiv.org/abs/1802.07273} {arXiv:1802.07273 [hep-ph]} \BibitemShut
  {NoStop}%
\bibitem [{\citenamefont {Liu}\ \emph {et~al.}(2019)\citenamefont {Liu},
  \citenamefont {Elwood}, \citenamefont {Evans},\ and\ \citenamefont
  {Thaler}}]{Liu:2018icu}%
  \BibitemOpen
  \bibfield  {author} {\bibinfo {author} {\bibfnamefont {H.}~\bibnamefont
  {Liu}}, \bibinfo {author} {\bibfnamefont {B.~D.}\ \bibnamefont {Elwood}},
  \bibinfo {author} {\bibfnamefont {M.}~\bibnamefont {Evans}}, \ and\ \bibinfo
  {author} {\bibfnamefont {J.}~\bibnamefont {Thaler}},\ }\href {\doibase
  10.1103/PhysRevD.100.023548} {\bibfield  {journal} {\bibinfo  {journal}
  {Phys. Rev. D}\ }\textbf {\bibinfo {volume} {100}},\ \bibinfo {pages}
  {023548} (\bibinfo {year} {2019})},\ \Eprint
  {http://arxiv.org/abs/1809.01656} {arXiv:1809.01656 [hep-ph]} \BibitemShut
  {NoStop}%
\bibitem [{\citenamefont {Obata}\ \emph {et~al.}(2018)\citenamefont {Obata},
  \citenamefont {Fujita},\ and\ \citenamefont {Michimura}}]{Obata:2018vvr}%
  \BibitemOpen
  \bibfield  {author} {\bibinfo {author} {\bibfnamefont {I.}~\bibnamefont
  {Obata}}, \bibinfo {author} {\bibfnamefont {T.}~\bibnamefont {Fujita}}, \
  and\ \bibinfo {author} {\bibfnamefont {Y.}~\bibnamefont {Michimura}},\ }\href
  {\doibase 10.1103/PhysRevLett.121.161301} {\bibfield  {journal} {\bibinfo
  {journal} {Phys. Rev. Lett.}\ }\textbf {\bibinfo {volume} {121}},\ \bibinfo
  {pages} {161301} (\bibinfo {year} {2018})},\ \Eprint
  {http://arxiv.org/abs/1805.11753} {arXiv:1805.11753 [astro-ph.CO]}
  \BibitemShut {NoStop}%
\bibitem [{\citenamefont {Feng}\ \emph {et~al.}(2018)\citenamefont {Feng},
  \citenamefont {Galon}, \citenamefont {Kling},\ and\ \citenamefont
  {Trojanowski}}]{Feng:2018noy}%
  \BibitemOpen
  \bibfield  {author} {\bibinfo {author} {\bibfnamefont {J.~L.}\ \bibnamefont
  {Feng}}, \bibinfo {author} {\bibfnamefont {I.}~\bibnamefont {Galon}},
  \bibinfo {author} {\bibfnamefont {F.}~\bibnamefont {Kling}}, \ and\ \bibinfo
  {author} {\bibfnamefont {S.}~\bibnamefont {Trojanowski}},\ }\href {\doibase
  10.1103/PhysRevD.98.055021} {\bibfield  {journal} {\bibinfo  {journal} {Phys.
  Rev. D}\ }\textbf {\bibinfo {volume} {98}},\ \bibinfo {pages} {055021}
  (\bibinfo {year} {2018})},\ \Eprint {http://arxiv.org/abs/1806.02348}
  {arXiv:1806.02348 [hep-ph]} \BibitemShut {NoStop}%
\bibitem [{\citenamefont {Berlin}\ \emph {et~al.}(2019)\citenamefont {Berlin},
  \citenamefont {Blinov}, \citenamefont {Krnjaic}, \citenamefont {Schuster},\
  and\ \citenamefont {Toro}}]{Berlin:2018bsc}%
  \BibitemOpen
  \bibfield  {author} {\bibinfo {author} {\bibfnamefont {A.}~\bibnamefont
  {Berlin}}, \bibinfo {author} {\bibfnamefont {N.}~\bibnamefont {Blinov}},
  \bibinfo {author} {\bibfnamefont {G.}~\bibnamefont {Krnjaic}}, \bibinfo
  {author} {\bibfnamefont {P.}~\bibnamefont {Schuster}}, \ and\ \bibinfo
  {author} {\bibfnamefont {N.}~\bibnamefont {Toro}},\ }\href {\doibase
  10.1103/PhysRevD.99.075001} {\bibfield  {journal} {\bibinfo  {journal} {Phys.
  Rev. D}\ }\textbf {\bibinfo {volume} {99}},\ \bibinfo {pages} {075001}
  (\bibinfo {year} {2019})},\ \Eprint {http://arxiv.org/abs/1807.01730}
  {arXiv:1807.01730 [hep-ph]} \BibitemShut {NoStop}%
\bibitem [{\citenamefont {Åkesson}\ \emph {et~al.}(2018)\citenamefont
  {Åkesson} \emph {et~al.}}]{Akesson:2018vlm}%
  \BibitemOpen
  \bibfield  {author} {\bibinfo {author} {\bibfnamefont {T.}~\bibnamefont
  {Åkesson}} \emph {et~al.} (\bibinfo {collaboration} {LDMX}),\ }\href@noop {}
  {\  (\bibinfo {year} {2018})},\ \Eprint {http://arxiv.org/abs/1808.05219}
  {arXiv:1808.05219 [hep-ex]} \BibitemShut {NoStop}%
\bibitem [{\citenamefont {Volpe}(2019)}]{Volpe:2019nzt}%
  \BibitemOpen
  \bibfield  {author} {\bibinfo {author} {\bibfnamefont {R.}~\bibnamefont
  {Volpe}},\ }in\ \href@noop {} {\emph {\bibinfo {booktitle} {{Meeting of the
  Division of Particles and Fields of the American Physical Society (DPF2019)
  Boston, Massachusetts, July 29-August 2, 2019}}}}\ (\bibinfo {year} {2019})\
  \Eprint {http://arxiv.org/abs/1910.10429} {arXiv:1910.10429 [hep-ex]}
  \BibitemShut {NoStop}%
\bibitem [{\citenamefont {Berlin}\ \emph {et~al.}(2018)\citenamefont {Berlin},
  \citenamefont {Gori}, \citenamefont {Schuster},\ and\ \citenamefont
  {Toro}}]{Berlin:2018pwi}%
  \BibitemOpen
  \bibfield  {author} {\bibinfo {author} {\bibfnamefont {A.}~\bibnamefont
  {Berlin}}, \bibinfo {author} {\bibfnamefont {S.}~\bibnamefont {Gori}},
  \bibinfo {author} {\bibfnamefont {P.}~\bibnamefont {Schuster}}, \ and\
  \bibinfo {author} {\bibfnamefont {N.}~\bibnamefont {Toro}},\ }\href {\doibase
  10.1103/PhysRevD.98.035011} {\bibfield  {journal} {\bibinfo  {journal} {Phys.
  Rev. D}\ }\textbf {\bibinfo {volume} {98}},\ \bibinfo {pages} {035011}
  (\bibinfo {year} {2018})},\ \Eprint {http://arxiv.org/abs/1804.00661}
  {arXiv:1804.00661 [hep-ph]} \BibitemShut {NoStop}%
\bibitem [{\citenamefont {Alekhin}\ \emph {et~al.}(2016)\citenamefont {Alekhin}
  \emph {et~al.}}]{Alekhin:2015byh}%
  \BibitemOpen
  \bibfield  {author} {\bibinfo {author} {\bibfnamefont {S.}~\bibnamefont
  {Alekhin}} \emph {et~al.},\ }\href {\doibase 10.1088/0034-4885/79/12/124201}
  {\bibfield  {journal} {\bibinfo  {journal} {Rept. Prog. Phys.}\ }\textbf
  {\bibinfo {volume} {79}},\ \bibinfo {pages} {124201} (\bibinfo {year}
  {2016})},\ \Eprint {http://arxiv.org/abs/1504.04855} {arXiv:1504.04855
  [hep-ph]} \BibitemShut {NoStop}%
\bibitem [{\citenamefont {Bauer}\ \emph {et~al.}(2019)\citenamefont {Bauer},
  \citenamefont {Heiles}, \citenamefont {Neubert},\ and\ \citenamefont
  {Thamm}}]{Bauer:2018uxu}%
  \BibitemOpen
  \bibfield  {author} {\bibinfo {author} {\bibfnamefont {M.}~\bibnamefont
  {Bauer}}, \bibinfo {author} {\bibfnamefont {M.}~\bibnamefont {Heiles}},
  \bibinfo {author} {\bibfnamefont {M.}~\bibnamefont {Neubert}}, \ and\
  \bibinfo {author} {\bibfnamefont {A.}~\bibnamefont {Thamm}},\ }\href
  {\doibase 10.1140/epjc/s10052-019-6587-9} {\bibfield  {journal} {\bibinfo
  {journal} {Eur. Phys. J. C}\ }\textbf {\bibinfo {volume} {79}},\ \bibinfo
  {pages} {74} (\bibinfo {year} {2019})},\ \Eprint
  {http://arxiv.org/abs/1808.10323} {arXiv:1808.10323 [hep-ph]} \BibitemShut
  {NoStop}%
\bibitem [{\citenamefont {Astier}\ \emph {et~al.}(2000)\citenamefont {Astier}
  \emph {et~al.}}]{Astier:2000gx}%
  \BibitemOpen
  \bibfield  {author} {\bibinfo {author} {\bibfnamefont {P.}~\bibnamefont
  {Astier}} \emph {et~al.} (\bibinfo {collaboration} {NOMAD}),\ }\href
  {\doibase 10.1016/S0370-2693(00)00375-0} {\bibfield  {journal} {\bibinfo
  {journal} {Phys. Lett. B}\ }\textbf {\bibinfo {volume} {479}},\ \bibinfo
  {pages} {371} (\bibinfo {year} {2000})}\BibitemShut {NoStop}%
\bibitem [{\citenamefont {Bonivento}\ \emph {et~al.}(2019)\citenamefont
  {Bonivento}, \citenamefont {Kim},\ and\ \citenamefont
  {Sinha}}]{Bonivento:2019sri}%
  \BibitemOpen
  \bibfield  {author} {\bibinfo {author} {\bibfnamefont {W.~M.}\ \bibnamefont
  {Bonivento}}, \bibinfo {author} {\bibfnamefont {D.}~\bibnamefont {Kim}}, \
  and\ \bibinfo {author} {\bibfnamefont {K.}~\bibnamefont {Sinha}},\
  }\href@noop {} {\  (\bibinfo {year} {2019})},\ \Eprint
  {http://arxiv.org/abs/1909.03071} {arXiv:1909.03071 [hep-ph]} \BibitemShut
  {NoStop}%
\bibitem [{\citenamefont {Abe}\ \emph {et~al.}(2013)\citenamefont {Abe} \emph
  {et~al.}}]{Abe:2012ut}%
  \BibitemOpen
  \bibfield  {author} {\bibinfo {author} {\bibfnamefont {K.}~\bibnamefont
  {Abe}} \emph {et~al.},\ }\href {\doibase 10.1016/j.physletb.2013.05.060}
  {\bibfield  {journal} {\bibinfo  {journal} {Phys. Lett. B}\ }\textbf
  {\bibinfo {volume} {724}},\ \bibinfo {pages} {46} (\bibinfo {year} {2013})},\
  \Eprint {http://arxiv.org/abs/1212.6153} {arXiv:1212.6153 [astro-ph.CO]}
  \BibitemShut {NoStop}%
\bibitem [{\citenamefont {Armengaud}\ \emph {et~al.}(2018)\citenamefont
  {Armengaud} \emph {et~al.}}]{Armengaud:2018cuy}%
  \BibitemOpen
  \bibfield  {author} {\bibinfo {author} {\bibfnamefont {E.}~\bibnamefont
  {Armengaud}} \emph {et~al.} (\bibinfo {collaboration} {EDELWEISS}),\ }\href
  {\doibase 10.1103/PhysRevD.98.082004} {\bibfield  {journal} {\bibinfo
  {journal} {Phys. Rev. D}\ }\textbf {\bibinfo {volume} {98}},\ \bibinfo
  {pages} {082004} (\bibinfo {year} {2018})},\ \Eprint
  {http://arxiv.org/abs/1808.02340} {arXiv:1808.02340 [hep-ex]} \BibitemShut
  {NoStop}%
\bibitem [{\citenamefont {Akerib}\ \emph {et~al.}(2017)\citenamefont {Akerib}
  \emph {et~al.}}]{Akerib:2017uem}%
  \BibitemOpen
  \bibfield  {author} {\bibinfo {author} {\bibfnamefont {D.~S.}\ \bibnamefont
  {Akerib}} \emph {et~al.} (\bibinfo {collaboration} {LUX}),\ }\href {\doibase
  10.1103/PhysRevLett.118.261301} {\bibfield  {journal} {\bibinfo  {journal}
  {Phys. Rev. Lett.}\ }\textbf {\bibinfo {volume} {118}},\ \bibinfo {pages}
  {261301} (\bibinfo {year} {2017})},\ \Eprint
  {http://arxiv.org/abs/1704.02297} {arXiv:1704.02297 [astro-ph.CO]}
  \BibitemShut {NoStop}%
\bibitem [{\citenamefont {Fu}\ \emph {et~al.}(2017)\citenamefont {Fu} \emph
  {et~al.}}]{Fu:2017lfc}%
  \BibitemOpen
  \bibfield  {author} {\bibinfo {author} {\bibfnamefont {C.}~\bibnamefont {Fu}}
  \emph {et~al.} (\bibinfo {collaboration} {PandaX}),\ }\href {\doibase
  10.1103/PhysRevLett.119.181806} {\bibfield  {journal} {\bibinfo  {journal}
  {Phys. Rev. Lett.}\ }\textbf {\bibinfo {volume} {119}},\ \bibinfo {pages}
  {181806} (\bibinfo {year} {2017})},\ \Eprint
  {http://arxiv.org/abs/1707.07921} {arXiv:1707.07921 [hep-ex]} \BibitemShut
  {NoStop}%
\bibitem [{\citenamefont {Aprile}\ \emph {et~al.}(2019)\citenamefont {Aprile}
  \emph {et~al.}}]{Aprile:2019xxb}%
  \BibitemOpen
  \bibfield  {author} {\bibinfo {author} {\bibfnamefont {E.}~\bibnamefont
  {Aprile}} \emph {et~al.} (\bibinfo {collaboration} {XENON}),\ }\href@noop {}
  {\  (\bibinfo {year} {2019})},\ \Eprint {http://arxiv.org/abs/1907.11485}
  {arXiv:1907.11485 [hep-ex]} \BibitemShut {NoStop}%
\bibitem [{\citenamefont {Aralis}\ \emph {et~al.}(2019)\citenamefont {Aralis}
  \emph {et~al.}}]{Aralis:2019nfa}%
  \BibitemOpen
  \bibfield  {author} {\bibinfo {author} {\bibfnamefont {T.}~\bibnamefont
  {Aralis}} \emph {et~al.} (\bibinfo {collaboration} {SuperCDMS}),\ }\href@noop
  {} {\  (\bibinfo {year} {2019})},\ \Eprint {http://arxiv.org/abs/1911.11905}
  {arXiv:1911.11905 [hep-ex]} \BibitemShut {NoStop}%
\bibitem [{\citenamefont {Davoudiasl}\ and\ \citenamefont
  {Huber}(2009)}]{Davoudiasl:2009fe}%
  \BibitemOpen
  \bibfield  {author} {\bibinfo {author} {\bibfnamefont {H.}~\bibnamefont
  {Davoudiasl}}\ and\ \bibinfo {author} {\bibfnamefont {P.}~\bibnamefont
  {Huber}},\ }\href {\doibase 10.1103/PhysRevD.79.095024} {\bibfield  {journal}
  {\bibinfo  {journal} {Phys. Rev. D}\ }\textbf {\bibinfo {volume} {79}},\
  \bibinfo {pages} {095024} (\bibinfo {year} {2009})},\ \Eprint
  {http://arxiv.org/abs/0903.0618} {arXiv:0903.0618 [hep-ph]} \BibitemShut
  {NoStop}%
\bibitem [{\citenamefont {Bernabei}\ \emph {et~al.}(2001)\citenamefont
  {Bernabei} \emph {et~al.}}]{Bernabei:2001ny}%
  \BibitemOpen
  \bibfield  {author} {\bibinfo {author} {\bibfnamefont {R.}~\bibnamefont
  {Bernabei}} \emph {et~al.},\ }\href {\doibase 10.1016/S0370-2693(01)00840-1}
  {\bibfield  {journal} {\bibinfo  {journal} {Phys.\ Lett.\ B}\ }\textbf
  {\bibinfo {volume} {515}},\ \bibinfo {pages} {6} (\bibinfo {year}
  {2001})}\BibitemShut {NoStop}%
\bibitem [{\citenamefont {Armengaud}\ \emph {et~al.}(2013)\citenamefont
  {Armengaud} \emph {et~al.}}]{Armengaud:2013rta}%
  \BibitemOpen
  \bibfield  {author} {\bibinfo {author} {\bibfnamefont {E.}~\bibnamefont
  {Armengaud}} \emph {et~al.},\ }\href {\doibase 10.1088/1475-7516/2013/11/067}
  {\bibfield  {journal} {\bibinfo  {journal} {JCAP}\ }\textbf {\bibinfo
  {volume} {11}},\ \bibinfo {pages} {067} (\bibinfo {year} {2013})},\ \Eprint
  {http://arxiv.org/abs/1307.1488} {arXiv:1307.1488 [astro-ph.CO]} \BibitemShut
  {NoStop}%
\bibitem [{\citenamefont {Oka}\ \emph {et~al.}(2017)\citenamefont {Oka} \emph
  {et~al.}}]{Oka:2017rnn}%
  \BibitemOpen
  \bibfield  {author} {\bibinfo {author} {\bibfnamefont {N.}~\bibnamefont
  {Oka}} \emph {et~al.} (\bibinfo {collaboration} {XMASS}),\ }\href {\doibase
  10.1093/ptep/ptx137} {\bibfield  {journal} {\bibinfo  {journal} {PTEP}\
  }\textbf {\bibinfo {volume} {2017}},\ \bibinfo {pages} {103C01} (\bibinfo
  {year} {2017})},\ \Eprint {http://arxiv.org/abs/1707.08995} {arXiv:1707.08995
  [hep-ex]} \BibitemShut {NoStop}%
\bibitem [{\citenamefont {Moriyama}(1995)}]{Moriyama:1995bz}%
  \BibitemOpen
  \bibfield  {author} {\bibinfo {author} {\bibfnamefont {S.}~\bibnamefont
  {Moriyama}},\ }\href {\doibase 10.1103/PhysRevLett.75.3222} {\bibfield
  {journal} {\bibinfo  {journal} {Phys.\ Rev.\ Lett.}\ }\textbf {\bibinfo
  {volume} {75}},\ \bibinfo {pages} {3222} (\bibinfo {year} {1995})},\ \Eprint
  {http://arxiv.org/abs/hep-ph/9504318} {arXiv:hep-ph/9504318} \BibitemShut
  {NoStop}%
\bibitem [{\citenamefont {Krcmar}\ \emph {et~al.}(1998)\citenamefont {Krcmar},
  \citenamefont {Krecak}, \citenamefont {Stipcevic}, \citenamefont {Ljubicic},\
  and\ \citenamefont {Bradley}}]{Krcmar:1998xn}%
  \BibitemOpen
  \bibfield  {author} {\bibinfo {author} {\bibfnamefont {M.}~\bibnamefont
  {Krcmar}}, \bibinfo {author} {\bibfnamefont {Z.}~\bibnamefont {Krecak}},
  \bibinfo {author} {\bibfnamefont {M.}~\bibnamefont {Stipcevic}}, \bibinfo
  {author} {\bibfnamefont {A.}~\bibnamefont {Ljubicic}}, \ and\ \bibinfo
  {author} {\bibfnamefont {D.}~\bibnamefont {Bradley}},\ }\href {\doibase
  10.1016/S0370-2693(98)01231-3} {\bibfield  {journal} {\bibinfo  {journal}
  {Phys.\ Lett.\ B}\ }\textbf {\bibinfo {volume} {442}},\ \bibinfo {pages} {38}
  (\bibinfo {year} {1998})},\ \Eprint {http://arxiv.org/abs/nucl-ex/9801005}
  {arXiv:nucl-ex/9801005} \BibitemShut {NoStop}%
\bibitem [{\citenamefont {Krcmar}\ \emph {et~al.}(2001)\citenamefont {Krcmar},
  \citenamefont {Krecak}, \citenamefont {Ljubicic}, \citenamefont {Stipcevic},\
  and\ \citenamefont {Bradley}}]{Krcmar:2001si}%
  \BibitemOpen
  \bibfield  {author} {\bibinfo {author} {\bibfnamefont {M.}~\bibnamefont
  {Krcmar}}, \bibinfo {author} {\bibfnamefont {Z.}~\bibnamefont {Krecak}},
  \bibinfo {author} {\bibfnamefont {A.}~\bibnamefont {Ljubicic}}, \bibinfo
  {author} {\bibfnamefont {M.}~\bibnamefont {Stipcevic}}, \ and\ \bibinfo
  {author} {\bibfnamefont {D.}~\bibnamefont {Bradley}},\ }\href {\doibase
  10.1103/PhysRevD.64.115016} {\bibfield  {journal} {\bibinfo  {journal}
  {Phys.\ Rev.\ D}\ }\textbf {\bibinfo {volume} {64}},\ \bibinfo {pages}
  {115016} (\bibinfo {year} {2001})},\ \Eprint
  {http://arxiv.org/abs/hep-ex/0104035} {arXiv:hep-ex/0104035} \BibitemShut
  {NoStop}%
\bibitem [{\citenamefont {Derbin}\ \emph {et~al.}(2009)\citenamefont {Derbin},
  \citenamefont {Bakhlanov}, \citenamefont {Egorov}, \citenamefont
  {Mitropolsky}, \citenamefont {Muratova}, \citenamefont {Semenov},\ and\
  \citenamefont {Unzhakov}}]{Derbin:2009jw}%
  \BibitemOpen
  \bibfield  {author} {\bibinfo {author} {\bibfnamefont {A.}~\bibnamefont
  {Derbin}}, \bibinfo {author} {\bibfnamefont {S.}~\bibnamefont {Bakhlanov}},
  \bibinfo {author} {\bibfnamefont {A.}~\bibnamefont {Egorov}}, \bibinfo
  {author} {\bibfnamefont {I.}~\bibnamefont {Mitropolsky}}, \bibinfo {author}
  {\bibfnamefont {V.}~\bibnamefont {Muratova}}, \bibinfo {author}
  {\bibfnamefont {D.}~\bibnamefont {Semenov}}, \ and\ \bibinfo {author}
  {\bibfnamefont {E.}~\bibnamefont {Unzhakov}},\ }\href {\doibase
  10.1016/j.physletb.2009.06.016} {\bibfield  {journal} {\bibinfo  {journal}
  {Phys.\ Lett.\ B}\ }\textbf {\bibinfo {volume} {678}},\ \bibinfo {pages}
  {181} (\bibinfo {year} {2009})},\ \Eprint {http://arxiv.org/abs/0904.3443}
  {arXiv:0904.3443 [hep-ph]} \BibitemShut {NoStop}%
\bibitem [{\citenamefont {Gavrilyuk}\ \emph {et~al.}(2018)\citenamefont
  {Gavrilyuk} \emph {et~al.}}]{Gavrilyuk:2018jdi}%
  \BibitemOpen
  \bibfield  {author} {\bibinfo {author} {\bibfnamefont {Y.}~\bibnamefont
  {Gavrilyuk}} \emph {et~al.},\ }\href {\doibase 10.1134/S1063779618010136}
  {\bibfield  {journal} {\bibinfo  {journal} {Phys.\ Part.\ Nucl.}\ }\textbf
  {\bibinfo {volume} {49}},\ \bibinfo {pages} {94} (\bibinfo {year}
  {2018})}\BibitemShut {NoStop}%
\bibitem [{\citenamefont {Creswick}\ \emph {et~al.}(2018)\citenamefont
  {Creswick}, \citenamefont {Li}, \citenamefont {Avignone},\ and\ \citenamefont
  {Wang}}]{Creswick:2018stb}%
  \BibitemOpen
  \bibfield  {author} {\bibinfo {author} {\bibfnamefont {R.}~\bibnamefont
  {Creswick}}, \bibinfo {author} {\bibfnamefont {D.}~\bibnamefont {Li}},
  \bibinfo {author} {\bibfnamefont {F.~T.}\ \bibnamefont {Avignone}}, \ and\
  \bibinfo {author} {\bibfnamefont {Y.}~\bibnamefont {Wang}},\ }in\ \href
  {\doibase 10.3204/DESY-PROC-2017-02/creswick\_richard} {\emph {\bibinfo
  {booktitle} {{Proceedings, 13th Patras Workshop on Axions, WIMPs and WISPs,
  (PATRAS 2017)}: {Thessaloniki, Greece, 15 May 2017 - 19, 2017}}}}\ (\bibinfo
  {year} {2018})\ pp.\ \bibinfo {pages} {11--14}\BibitemShut {NoStop}%
\bibitem [{\citenamefont {Sikivie}(2020)}]{Sikivie:2020zpn}%
  \BibitemOpen
  \bibfield  {author} {\bibinfo {author} {\bibfnamefont {P.}~\bibnamefont
  {Sikivie}},\ }\href@noop {} {\  (\bibinfo {year} {2020})},\ \Eprint
  {http://arxiv.org/abs/2003.02206} {arXiv:2003.02206 [hep-ph]} \BibitemShut
  {NoStop}%
\bibitem [{\citenamefont {Agnolet}\ \emph {et~al.}(2017)\citenamefont {Agnolet}
  \emph {et~al.}}]{Agnolet:2016zir}%
  \BibitemOpen
  \bibfield  {author} {\bibinfo {author} {\bibfnamefont {G.}~\bibnamefont
  {Agnolet}} \emph {et~al.} (\bibinfo {collaboration} {MINER}),\ }\href
  {\doibase 10.1016/j.nima.2017.02.024} {\bibfield  {journal} {\bibinfo
  {journal} {Nucl. Instrum. Meth.}\ }\textbf {\bibinfo {volume} {A853}},\
  \bibinfo {pages} {53} (\bibinfo {year} {2017})},\ \Eprint
  {http://arxiv.org/abs/1609.02066} {arXiv:1609.02066 [physics.ins-det]}
  \BibitemShut {NoStop}%
\bibitem [{\citenamefont {Aguilar-Arevalo}\ \emph {et~al.}(2016)\citenamefont
  {Aguilar-Arevalo} \emph {et~al.}}]{Aguilar-Arevalo:2016khx}%
  \BibitemOpen
  \bibfield  {author} {\bibinfo {author} {\bibfnamefont {A.}~\bibnamefont
  {Aguilar-Arevalo}} \emph {et~al.} (\bibinfo {collaboration} {CONNIE}),\
  }\bibfield  {booktitle} {\emph {\bibinfo {booktitle} {{Proceedings, 15th
  Mexican Workshop on Particles and Fields (MWPF 2015): Mazatlán, México,
  November 2-6, 2015}}},\ }\href {\doibase 10.1088/1742-6596/761/1/012057}
  {\bibfield  {journal} {\bibinfo  {journal} {J. Phys. Conf. Ser.}\ }\textbf
  {\bibinfo {volume} {761}},\ \bibinfo {pages} {012057} (\bibinfo {year}
  {2016})},\ \Eprint {http://arxiv.org/abs/1608.01565} {arXiv:1608.01565
  [physics.ins-det]} \BibitemShut {NoStop}%
\bibitem [{\citenamefont {Buck}\ \emph {et~al.}(2020)\citenamefont {Buck} \emph
  {et~al.}}]{Buck:2020opf}%
  \BibitemOpen
  \bibfield  {author} {\bibinfo {author} {\bibfnamefont {C.}~\bibnamefont
  {Buck}} \emph {et~al.},\ }\bibfield  {booktitle} {\emph {\bibinfo {booktitle}
  {{Proceedings, 15th International Conference on Topics in Astroparticle and
  Underground Physics (TAUP 2017): Sudbury, Ontario, Canada, July 24-28,
  2017}}},\ }\href {\doibase 10.1088/1742-6596/1342/1/012094} {\bibfield
  {journal} {\bibinfo  {journal} {J. Phys. Conf. Ser.}\ }\textbf {\bibinfo
  {volume} {1342}},\ \bibinfo {pages} {012094} (\bibinfo {year}
  {2020})}\BibitemShut {NoStop}%
\bibitem [{\citenamefont {Strauss}\ \emph {et~al.}(2017)\citenamefont {Strauss}
  \emph {et~al.}}]{Strauss:2017cuu}%
  \BibitemOpen
  \bibfield  {author} {\bibinfo {author} {\bibfnamefont {R.}~\bibnamefont
  {Strauss}} \emph {et~al.},\ }\href {\doibase 10.1140/epjc/s10052-017-5068-2}
  {\bibfield  {journal} {\bibinfo  {journal} {Eur. Phys. J. C}\ }\textbf
  {\bibinfo {volume} {77}},\ \bibinfo {pages} {506} (\bibinfo {year} {2017})},\
  \Eprint {http://arxiv.org/abs/1704.04320} {arXiv:1704.04320
  [physics.ins-det]} \BibitemShut {NoStop}%
\bibitem [{\citenamefont {Chang}\ \emph {et~al.}(2007)\citenamefont {Chang}
  \emph {et~al.}}]{Chang:2006ug}%
  \BibitemOpen
  \bibfield  {author} {\bibinfo {author} {\bibfnamefont {H.~M.}\ \bibnamefont
  {Chang}} \emph {et~al.} (\bibinfo {collaboration} {TEXONO}),\ }\href
  {\doibase 10.1103/PhysRevD.75.052004} {\bibfield  {journal} {\bibinfo
  {journal} {Phys. Rev. D}\ }\textbf {\bibinfo {volume} {75}},\ \bibinfo
  {pages} {052004} (\bibinfo {year} {2007})},\ \Eprint
  {http://arxiv.org/abs/hep-ex/0609001} {arXiv:hep-ex/0609001 [hep-ex]}
  \BibitemShut {NoStop}%
\bibitem [{\citenamefont {Jaeckel}\ \emph {et~al.}(2007)\citenamefont
  {Jaeckel}, \citenamefont {Masso}, \citenamefont {Redondo}, \citenamefont
  {Ringwald},\ and\ \citenamefont {Takahashi}}]{Jaeckel:2006xm}%
  \BibitemOpen
  \bibfield  {author} {\bibinfo {author} {\bibfnamefont {J.}~\bibnamefont
  {Jaeckel}}, \bibinfo {author} {\bibfnamefont {E.}~\bibnamefont {Masso}},
  \bibinfo {author} {\bibfnamefont {J.}~\bibnamefont {Redondo}}, \bibinfo
  {author} {\bibfnamefont {A.}~\bibnamefont {Ringwald}}, \ and\ \bibinfo
  {author} {\bibfnamefont {F.}~\bibnamefont {Takahashi}},\ }\href {\doibase
  10.1103/PhysRevD.75.013004} {\bibfield  {journal} {\bibinfo  {journal} {Phys.
  Rev. D}\ }\textbf {\bibinfo {volume} {75}},\ \bibinfo {pages} {013004}
  (\bibinfo {year} {2007})},\ \Eprint {http://arxiv.org/abs/hep-ph/0610203}
  {arXiv:hep-ph/0610203 [hep-ph]} \BibitemShut {NoStop}%
\bibitem [{\citenamefont {Khoury}\ and\ \citenamefont
  {Weltman}(2004)}]{Khoury:2003aq}%
  \BibitemOpen
  \bibfield  {author} {\bibinfo {author} {\bibfnamefont {J.}~\bibnamefont
  {Khoury}}\ and\ \bibinfo {author} {\bibfnamefont {A.}~\bibnamefont
  {Weltman}},\ }\href {\doibase 10.1103/PhysRevLett.93.171104} {\bibfield
  {journal} {\bibinfo  {journal} {Phys. Rev. Lett.}\ }\textbf {\bibinfo
  {volume} {93}},\ \bibinfo {pages} {171104} (\bibinfo {year} {2004})},\
  \Eprint {http://arxiv.org/abs/astro-ph/0309300} {arXiv:astro-ph/0309300
  [astro-ph]} \BibitemShut {NoStop}%
\bibitem [{\citenamefont {Masso}\ and\ \citenamefont
  {Redondo}(2005)}]{Masso:2005ym}%
  \BibitemOpen
  \bibfield  {author} {\bibinfo {author} {\bibfnamefont {E.}~\bibnamefont
  {Masso}}\ and\ \bibinfo {author} {\bibfnamefont {J.}~\bibnamefont
  {Redondo}},\ }\href {\doibase 10.1088/1475-7516/2005/09/015} {\bibfield
  {journal} {\bibinfo  {journal} {JCAP}\ }\textbf {\bibinfo {volume} {0509}},\
  \bibinfo {pages} {015} (\bibinfo {year} {2005})},\ \Eprint
  {http://arxiv.org/abs/hep-ph/0504202} {arXiv:hep-ph/0504202 [hep-ph]}
  \BibitemShut {NoStop}%
\bibitem [{\citenamefont {Masso}\ and\ \citenamefont
  {Redondo}(2006)}]{Masso:2006gc}%
  \BibitemOpen
  \bibfield  {author} {\bibinfo {author} {\bibfnamefont {E.}~\bibnamefont
  {Masso}}\ and\ \bibinfo {author} {\bibfnamefont {J.}~\bibnamefont
  {Redondo}},\ }\href {\doibase 10.1103/PhysRevLett.97.151802} {\bibfield
  {journal} {\bibinfo  {journal} {Phys. Rev. Lett.}\ }\textbf {\bibinfo
  {volume} {97}},\ \bibinfo {pages} {151802} (\bibinfo {year} {2006})},\
  \Eprint {http://arxiv.org/abs/hep-ph/0606163} {arXiv:hep-ph/0606163 [hep-ph]}
  \BibitemShut {NoStop}%
\bibitem [{\citenamefont {Dupays}\ \emph {et~al.}(2007)\citenamefont {Dupays},
  \citenamefont {Masso}, \citenamefont {Redondo},\ and\ \citenamefont
  {Rizzo}}]{Dupays:2006dp}%
  \BibitemOpen
  \bibfield  {author} {\bibinfo {author} {\bibfnamefont {A.}~\bibnamefont
  {Dupays}}, \bibinfo {author} {\bibfnamefont {E.}~\bibnamefont {Masso}},
  \bibinfo {author} {\bibfnamefont {J.}~\bibnamefont {Redondo}}, \ and\
  \bibinfo {author} {\bibfnamefont {C.}~\bibnamefont {Rizzo}},\ }\href
  {\doibase 10.1103/PhysRevLett.98.131802} {\bibfield  {journal} {\bibinfo
  {journal} {Phys. Rev. Lett.}\ }\textbf {\bibinfo {volume} {98}},\ \bibinfo
  {pages} {131802} (\bibinfo {year} {2007})},\ \Eprint
  {http://arxiv.org/abs/hep-ph/0610286} {arXiv:hep-ph/0610286 [hep-ph]}
  \BibitemShut {NoStop}%
\bibitem [{\citenamefont {Mohapatra}\ and\ \citenamefont
  {Nasri}(2007)}]{Mohapatra:2006pv}%
  \BibitemOpen
  \bibfield  {author} {\bibinfo {author} {\bibfnamefont {R.~N.}\ \bibnamefont
  {Mohapatra}}\ and\ \bibinfo {author} {\bibfnamefont {S.}~\bibnamefont
  {Nasri}},\ }\href {\doibase 10.1103/PhysRevLett.98.050402} {\bibfield
  {journal} {\bibinfo  {journal} {Phys. Rev. Lett.}\ }\textbf {\bibinfo
  {volume} {98}},\ \bibinfo {pages} {050402} (\bibinfo {year} {2007})},\
  \Eprint {http://arxiv.org/abs/hep-ph/0610068} {arXiv:hep-ph/0610068 [hep-ph]}
  \BibitemShut {NoStop}%
\bibitem [{\citenamefont {Brax}\ \emph {et~al.}(2007)\citenamefont {Brax},
  \citenamefont {van~de Bruck},\ and\ \citenamefont {Davis}}]{Brax:2007ak}%
  \BibitemOpen
  \bibfield  {author} {\bibinfo {author} {\bibfnamefont {P.}~\bibnamefont
  {Brax}}, \bibinfo {author} {\bibfnamefont {C.}~\bibnamefont {van~de Bruck}},
  \ and\ \bibinfo {author} {\bibfnamefont {A.-C.}\ \bibnamefont {Davis}},\
  }\href {\doibase 10.1103/PhysRevLett.99.121103} {\bibfield  {journal}
  {\bibinfo  {journal} {Phys. Rev. Lett.}\ }\textbf {\bibinfo {volume} {99}},\
  \bibinfo {pages} {121103} (\bibinfo {year} {2007})},\ \Eprint
  {http://arxiv.org/abs/hep-ph/0703243} {arXiv:hep-ph/0703243 [HEP-PH]}
  \BibitemShut {NoStop}%
\bibitem [{\citenamefont {Primakoff}(1951)}]{Pirmakoff:1951pj}%
  \BibitemOpen
  \bibfield  {author} {\bibinfo {author} {\bibfnamefont {H.}~\bibnamefont
  {Primakoff}},\ }\href {\doibase 10.1103/PhysRev.81.899} {\bibfield  {journal}
  {\bibinfo  {journal} {Phys. Rev.}\ }\textbf {\bibinfo {volume} {81}},\
  \bibinfo {pages} {899} (\bibinfo {year} {1951})}\BibitemShut {NoStop}%
\bibitem [{\citenamefont {Tsai}(1986)}]{Tsai:1986tx}%
  \BibitemOpen
  \bibfield  {author} {\bibinfo {author} {\bibfnamefont {Y.-S.}\ \bibnamefont
  {Tsai}},\ }\bibfield  {booktitle} {\emph {\bibinfo {booktitle} {{Proceedings,
  23RD International Conference on High Energy Physics, JULY 16-23, 1986,
  Berkeley, CA}}},\ }\href {\doibase 10.1103/PhysRevD.34.1326} {\bibfield
  {journal} {\bibinfo  {journal} {Phys. Rev. D}\ }\textbf {\bibinfo {volume}
  {34}},\ \bibinfo {pages} {1326} (\bibinfo {year} {1986})}\BibitemShut
  {NoStop}%
\bibitem [{\citenamefont {Aloni}\ \emph {et~al.}(2019)\citenamefont {Aloni},
  \citenamefont {Fanelli}, \citenamefont {Soreq},\ and\ \citenamefont
  {Williams}}]{Aloni:2019ruo}%
  \BibitemOpen
  \bibfield  {author} {\bibinfo {author} {\bibfnamefont {D.}~\bibnamefont
  {Aloni}}, \bibinfo {author} {\bibfnamefont {C.}~\bibnamefont {Fanelli}},
  \bibinfo {author} {\bibfnamefont {Y.}~\bibnamefont {Soreq}}, \ and\ \bibinfo
  {author} {\bibfnamefont {M.}~\bibnamefont {Williams}},\ }\href {\doibase
  10.1103/PhysRevLett.123.071801} {\bibfield  {journal} {\bibinfo  {journal}
  {Phys. Rev. Lett.}\ }\textbf {\bibinfo {volume} {123}},\ \bibinfo {pages}
  {071801} (\bibinfo {year} {2019})},\ \Eprint
  {http://arxiv.org/abs/1903.03586} {arXiv:1903.03586 [hep-ph]} \BibitemShut
  {NoStop}%
\bibitem [{\citenamefont {Brodsky}\ \emph {et~al.}(1986)\citenamefont
  {Brodsky}, \citenamefont {Mottola}, \citenamefont {Muzinich},\ and\
  \citenamefont {Soldate}}]{Brodsky:1986mi}%
  \BibitemOpen
  \bibfield  {author} {\bibinfo {author} {\bibfnamefont {S.~J.}\ \bibnamefont
  {Brodsky}}, \bibinfo {author} {\bibfnamefont {E.}~\bibnamefont {Mottola}},
  \bibinfo {author} {\bibfnamefont {I.~J.}\ \bibnamefont {Muzinich}}, \ and\
  \bibinfo {author} {\bibfnamefont {M.}~\bibnamefont {Soldate}},\ }\href
  {\doibase 10.1103/PhysRevLett.56.1763, 10.1103/PhysRevLett.57.502.2}
  {\bibfield  {journal} {\bibinfo  {journal} {Phys. Rev. Lett.}\ }\textbf
  {\bibinfo {volume} {56}},\ \bibinfo {pages} {1763} (\bibinfo {year}
  {1986})},\ \bibinfo {note} {[Erratum: Phys. Rev.
  Lett.57,502(1986)]}\BibitemShut {NoStop}%
\bibitem [{\citenamefont {Chakrabarty}\ and\ \citenamefont
  {Jaeglé}(2019)}]{Chakrabarty:2019kdd}%
  \BibitemOpen
  \bibfield  {author} {\bibinfo {author} {\bibfnamefont {S.~S.}\ \bibnamefont
  {Chakrabarty}}\ and\ \bibinfo {author} {\bibfnamefont {I.}~\bibnamefont
  {Jaeglé}},\ }\href@noop {} {\  (\bibinfo {year} {2019})},\ \Eprint
  {http://arxiv.org/abs/1903.06225} {arXiv:1903.06225 [hep-ph]} \BibitemShut
  {NoStop}%
\bibitem [{\citenamefont {Gondolo}\ and\ \citenamefont
  {Raffelt}(2009)}]{Gondolo:2008dd}%
  \BibitemOpen
  \bibfield  {author} {\bibinfo {author} {\bibfnamefont {P.}~\bibnamefont
  {Gondolo}}\ and\ \bibinfo {author} {\bibfnamefont {G.~G.}\ \bibnamefont
  {Raffelt}},\ }\href {\doibase 10.1103/PhysRevD.79.107301} {\bibfield
  {journal} {\bibinfo  {journal} {Phys. Rev. D}\ }\textbf {\bibinfo {volume}
  {79}},\ \bibinfo {pages} {107301} (\bibinfo {year} {2009})},\ \Eprint
  {http://arxiv.org/abs/0807.2926} {arXiv:0807.2926 [astro-ph]} \BibitemShut
  {NoStop}%
\bibitem [{\citenamefont {Avignone}\ \emph {et~al.}(1988)\citenamefont
  {Avignone}, \citenamefont {Baktash}, \citenamefont {Barker}, \citenamefont
  {Calaprice}, \citenamefont {Dunford}, \citenamefont {Haxton}, \citenamefont
  {Kahana}, \citenamefont {Kouzes}, \citenamefont {Miley},\ and\ \citenamefont
  {Moltz}}]{Avignone:1988bv}%
  \BibitemOpen
  \bibfield  {author} {\bibinfo {author} {\bibfnamefont {F.~T.}\ \bibnamefont
  {Avignone}}, \bibinfo {author} {\bibfnamefont {C.}~\bibnamefont {Baktash}},
  \bibinfo {author} {\bibfnamefont {W.~C.}\ \bibnamefont {Barker}}, \bibinfo
  {author} {\bibfnamefont {F.~P.}\ \bibnamefont {Calaprice}}, \bibinfo {author}
  {\bibfnamefont {R.~W.}\ \bibnamefont {Dunford}}, \bibinfo {author}
  {\bibfnamefont {W.~C.}\ \bibnamefont {Haxton}}, \bibinfo {author}
  {\bibfnamefont {D.}~\bibnamefont {Kahana}}, \bibinfo {author} {\bibfnamefont
  {R.~T.}\ \bibnamefont {Kouzes}}, \bibinfo {author} {\bibfnamefont {H.~S.}\
  \bibnamefont {Miley}}, \ and\ \bibinfo {author} {\bibfnamefont {D.~M.}\
  \bibnamefont {Moltz}},\ }\href {\doibase 10.1103/PhysRevD.37.618} {\bibfield
  {journal} {\bibinfo  {journal} {Phys. Rev. D}\ }\textbf {\bibinfo {volume}
  {37}},\ \bibinfo {pages} {618} (\bibinfo {year} {1988})}\BibitemShut
  {NoStop}%
\bibitem [{\citenamefont {Bellini}\ \emph {et~al.}(2008)\citenamefont {Bellini}
  \emph {et~al.}}]{Bellini:2008zza}%
  \BibitemOpen
  \bibfield  {author} {\bibinfo {author} {\bibfnamefont {G.}~\bibnamefont
  {Bellini}} \emph {et~al.} (\bibinfo {collaboration} {Borexino}),\ }\href
  {\doibase 10.1140/epjc/s10052-008-0530-9} {\bibfield  {journal} {\bibinfo
  {journal} {Eur. Phys. J. C}\ }\textbf {\bibinfo {volume} {54}},\ \bibinfo
  {pages} {61} (\bibinfo {year} {2008})}\BibitemShut {NoStop}%
\bibitem [{\citenamefont {Berger}\ \emph {et~al.}(2010)\citenamefont {Berger}
  \emph {et~al.}}]{NIST:2019}%
  \BibitemOpen
  \bibfield  {author} {\bibinfo {author} {\bibfnamefont {M.}~\bibnamefont
  {Berger}} \emph {et~al.} (\bibinfo {collaboration} {NIST}),\ }\href {\doibase
  10.18434/T48G6X} {\enquote {\bibinfo {title} {{XCOM: Photon Cross Section
  Database (version 1.5)}},}\ } (\bibinfo {year} {2010})\BibitemShut {NoStop}%
\bibitem [{\citenamefont {Aristizabal~Sierra}\ \emph
  {et~al.}(2019)\citenamefont {Aristizabal~Sierra} \emph
  {et~al.}}]{AristizabalSierra:2019hcm}%
  \BibitemOpen
  \bibfield  {author} {\bibinfo {author} {\bibfnamefont {D.}~\bibnamefont
  {Aristizabal~Sierra}} \emph {et~al.},\ }\href {\doibase
  10.5281/zenodo.3489190} {\  (\bibinfo {year} {2019}),\
  10.5281/zenodo.3489190},\ \Eprint {http://arxiv.org/abs/1910.07450}
  {arXiv:1910.07450 [hep-ex]} \BibitemShut {NoStop}%
\end{thebibliography}%

\end{document}